\documentclass[12pt]{iopart}

 \expandafter\let\csname equation*\endcsname\relax
  \expandafter\let\csname endequation*\endcsname\relax

\usepackage{amsthm,amssymb,amsmath}
\usepackage{graphicx}
\usepackage{color}
\usepackage[framemethod=TikZ]{mdframed}
\usepackage{multirow}
\usepackage{mathptmx}
\usepackage{bm}
\usepackage{url}
\usepackage{tikz}
\usetikzlibrary{arrows,positioning,shapes.geometric}
\usepackage{graphicx}

\newcommand{\RR}{\mathbb{R}}

\newcommand{\NN}{\mathbb{N}}

\newcommand{\ours}{\texttt{DD-NLEM}}
\newcommand{\oursX}{\texttt{NLEM}}

\newcommand{\argmin}{\operatornamewithlimits{argmin}}

\newtheorem{thm}{Theorem}[section]

\newtheorem{Assumption}[thm]{Assumption}
\newtheorem{remark}[thm]{Remark}

\begin{document}

\title[f-wave extraction]{Single-lead f-wave extraction using diffusion geometry}
\author{John Malik${}^{1*}$, Neil Reed${}^{1*}$, Chun-Li Wang${}^{2,3\dagger}$, Hau-tieng Wu${}^{1,4\dagger}$}

\address{$^1$ Department of Mathematics, University of Toronto, Toronto, Ontario, Canada}
\address{$^2$ Cardiovascular Division, Department of Internal Medicine, Chang Gung Memorial Hospital, Linkou Medical Center, Taoyuan, Taiwan}
\address{$^3$ College of Medicine, Chang Gung University, Taoyuan, Taiwan}
\address{$^4$ Mathematics Division, National Center for Theoretical Sciences, Taipei, Taiwan}

\address{${}^*$: these two authors contribute equally to this work. ${}^\dagger$: co-correspondence.}
\ead{hauwu@math.toronto.edu}

\begin{abstract}
A novel single-lead f-wave extraction algorithm based on the modern diffusion geometry data analysis framework is proposed. The algorithm is essentially an averaged beat subtraction algorithm, where the ventricular activity template is estimated by combining a newly designed metric, the ``diffusion distance,'' and the non-local Euclidean median based on the non-linear manifold setup. {We coined the algorithm $\ours$. Two simulation schemes are considered, and the new algorithm $\ours$ outperforms traditional algorithms, including the average beat subtraction, principal component analysis, and adaptive singular value cancellation, in different evaluation metrics with statistical significance.} The clinical potential is shown in the real Holter signal, and we introduce a new score to evaluate the performance of the algorithm.
\end{abstract}


\pacs{,}
\submitto{\PM}
\maketitle

\section{Introduction}

Atrial fibrillation (Af) is the most commonly sustained arrhythmia encountered in clinical practice and continues to receive considerable research interest. Interventions such as rhythm or rate control improve Af-related symptoms and may preserve cardiac function. However, current Af management guidelines provide no treatment recommendations that take the various mechanisms and patterns of Af into account \cite{doi:10.1093/eurheartj/ehw210,January2246} and therefore tests are developed that quantify Af and guide its management. The fibrillation wave (f-wave) related analysis of the surface ECG or long-term Holter monitoring for Af patients is undoubtedly one of the most challenging questions encountered in the clinical practice \cite{Nattel2002,Bollmann2006};
for example, what is the mechanism underlying the initiation, termination, and maintenance of paroxysmal Af \cite{Julian2015}, and what is the outcome of Af treatment \cite{Lankveld2016}? A summary of the available information on f-wave analysis and its clinical applications can be found, for example, in \cite{Bollmann2006}.
The main step toward f-wave analysis via the surface ECG is decoupling the atrial activity (AA), specifically the f-wave, from the ventricular activity (VA) by so-called QRST cancellation. The main challenges are the broad spectral nature of the ventricular response, the frequency overlap of the f-wave and the VA, and the time-varying frequency and amplitude of the cardiac activity, etc \cite{Bollmann2006}.  

This challenge has attracted a lot of interest in the field, and there have been several algorithms proposed in the literature aiming to extract the f-wave from the surface ECG. These algorithms can be roughly classified into two types, depending on the number of ECG channels needed for the algorithm. The first type includes blind source separation algorithms like independent component analysis and principal component analysis (PCA) {\cite{Langley2000,Rieta2004a,Langley2006,Llinares2009,Donoso2013,Zeemering2014, CastellsRieta2005}}, spatiotemporal QRST cancellation \cite{Stridh2001,Lemay2005}, and  adaptive filtering with its variations \cite{ThakorZhu1991,Ciaccio2012,Petrenas2012,Vasquez2001}.
These approaches typically need multiple channels; they deal frequently with the standard 12-lead ECG signal or the body surface potential map \cite{Zeemering2014}. 
While these techniques have their own merits, it is well known that the performance is downgraded when we have only a few surface ECG channels available, since in this case, the spatial information of the cardiac activity is incomplete and unattainable.  Also, {12-lead recording} is not suitable for long-term monitoring {because it is inconvenient for the patient}. 

The second type includes algorithms designed for the situation when only single-lead ECG is available. The main idea beyond these cancellation algorithms is that the atrial activity is not ``synchronized'' to the ventricular activity. For example, the averaged beat subtraction (ABS) algorithm \cite{Slocum1985,Slocum1992,Holm1998,Shah2004} and its variations based on PCA, the singular value decomposition \cite{Castells2005,Alcaraz2008,Alcaraz:2011}, and the wavelet transform \cite{Senhadji2002} are available choices to remove the ventricular activity from the single-lead surface ECG. 
The main merit of single-lead f-wave analysis is its practicality for long-term monitoring. In most mobile health monitoring systems, the number of leads is less than three; in some cases, only one lead is available, e.g. in the long-term monitoring device of \cite{Barrett:2014}. 
There are two main steps in ABS-type algorithms. The first step is determining a pool of cardiac activities by a chosen metric, and the second step is determining the VA template associated with each cardiac activity. For the first step, most ABS-type algorithms take the temporal relationship as the metric, and a few consider morphological relationship, e.g. \cite[Section 3.3]{Alcaraz2008}. For the second step, taking the mean of all beats in the pool is the common approach, while researchers found that the principal component of the pool, like \cite{Castells2005,Alcaraz2008,Alcaraz:2011}, keeps more VA information and hence achieves a better f-wave recovery. 
In brief, the ABS-type algorithm could be viewed as a solution to the {\em single-lead blind source separation problem}.

From the signal processing viewpoint, there are at least two challenges faced by the single-lead f-wave extraction algorithm. The first challenge is caused by the inevitable non-stationary dynamics in a living system. In the ECG signal, these non-stationary dynamics arise as the time-varying morphology of the VA. In Af patients, the situation is further complicated by the frequently accompanying premature ventricular contractions (PVC). Thus, it is not easy to obtain a good template using consecutive VAs. On the other hand, due to the basic homeostasis assumption of the living system, we know that from time to time, the VA morphology should not be ``too different.'' Thus, while the VA morphology might not be the same in a set of consecutive cardiac activities, we could find several QRST complexes at different points in the recorded signal that have similar morphologies. 
However, it is not easy to find these morphologically similar QRST complexes due to the existence of the f-wave.  This second challenge is caused by the fact that the interference between the VA and the f-wave is complicated and exists both in the time domain and in the frequency domain. For a given QRST complex, a direct attempt to find other morphologically similar QRST complexes via any simple metric might fail. We need to design a metric that is insensitive to the f-wave but provides sufficient similarity information about the VA morphology. 
The third difficulty is the lack of ground truth. Note that while we know that the recorded ECG signal is composed of the f-wave and the VA, there is no ground truth about how the f-wave and the VA should behave, in general. The closest possibility, as far as the authors know, is collecting the intracardiac AA in the intracardiac electrophysiology study. However, the relationship between the intracardiac AA and the f-wave in the recorded surface ECG is not trivial. This leads to the difficulty of validating a proposed algorithm. To handle this difficulty, we need a set of simulated data sets that well-approximate real Af cases, and a new metric to evaluate how the algorithm performs in the real signal. 

While the existing single-lead f-wave extraction algorithms have their own merits and have been widely applied, the above-mentioned difficulties explain why extracted f-waves from single-lead recordings using the existing algorithms often contain large QRS residuals and distortions. These distorted f-waves cause unreliable clinical interference results. 
To sum up, having a systematic, reliable, and accurate method to extract the f-wave from a single-lead ECG recording for the purposes of monitoring, diagnosis and treatment is not only a challenging and interesting signal processing problem, but also an important task for clinical needs.

The main contribution of this work is proposing a novel single-lead f-wave extraction algorithm that is accurate and suitable for long-term monitoring. The proposed algorithm is of the ABS-type, and it takes the non-linear and non-Euclidean structure reflecting the physiological dynamics into account.  
In this algorithm, the two major steps in the ABS-type algorithms are systematically changed by taking the manifold learning framework into account. For the first step, based on the currently developed unsupervised machine learning algorithm, diffusion map (DM), we design a new metric, called diffusion distance (DD), to ``cluster'' cardiac activities. For the second step, we get the VA template by the non-local Euclidean median (NLEM) algorithm \cite{Chaudhury_Singer:2012} based on the non-linear geometric model, which is modeled to be a special manifold structure, the fiber bundle. 
Mathematically, the non-linear and non-stationary dynamics of the surface ECG signal are taken into account to construct a non-linear but low-dimensional structure for the ventricular activity, while the f-wave is viewed as the fiber. This model allows us to justify why the DD allows us to find all similar VAs, and how the NLEM algorithm gives us an accurate VA template. 
To validate the proposed algorithm, based on the existing literature and physiological knowledge, we design two simulations for the Af patients, and we propose a new metric, called {\em modified ventricular residue (mVR)}, to evaluate the performance of different algorithms. Finally, we apply the algorithm to the 24 hours Holter signal and show the final result.

The paper is organized in the following way. 
In Section \ref{Section:Algorithm}, we introduce the proposed single-lead f-wave extraction algorithm, and a new evaluation metric for real signals.
In Section \ref{Section:MaterialResult}, we validate our proposed algorithm on simulated signals as well as Holter signals. The simulated methods are detailed for the purpose of reproducibility.
We provide the discussion and conclusions in Section \ref{Section:Discussion}.
The theoretical details are postponed to \ref{Section:Model}, where we specify our model for the surface ECG signal composed of the f-wave and the VA, and the mathematical background of DM and DD.

\section{Method}\label{Section:Algorithm}

In this section, we describe the proposed single-lead f-wave extraction algorithm, the simulated database, and the metrics used to evaluate our algorithm. We call our algorithm $\ours$.  For readers interested in the theoretical background and argument, please see \ref{Section:Model}. For the purpose of reproducibility, the MATLAB code is available via request. 

\subsection{Diffusion-based f-wave extraction algorithm -- $\ours$}\label{subalgorithm}

\tikzstyle{line} = [draw, -latex']
\tikzstyle{arrow} = [thick,->,>=stealth]
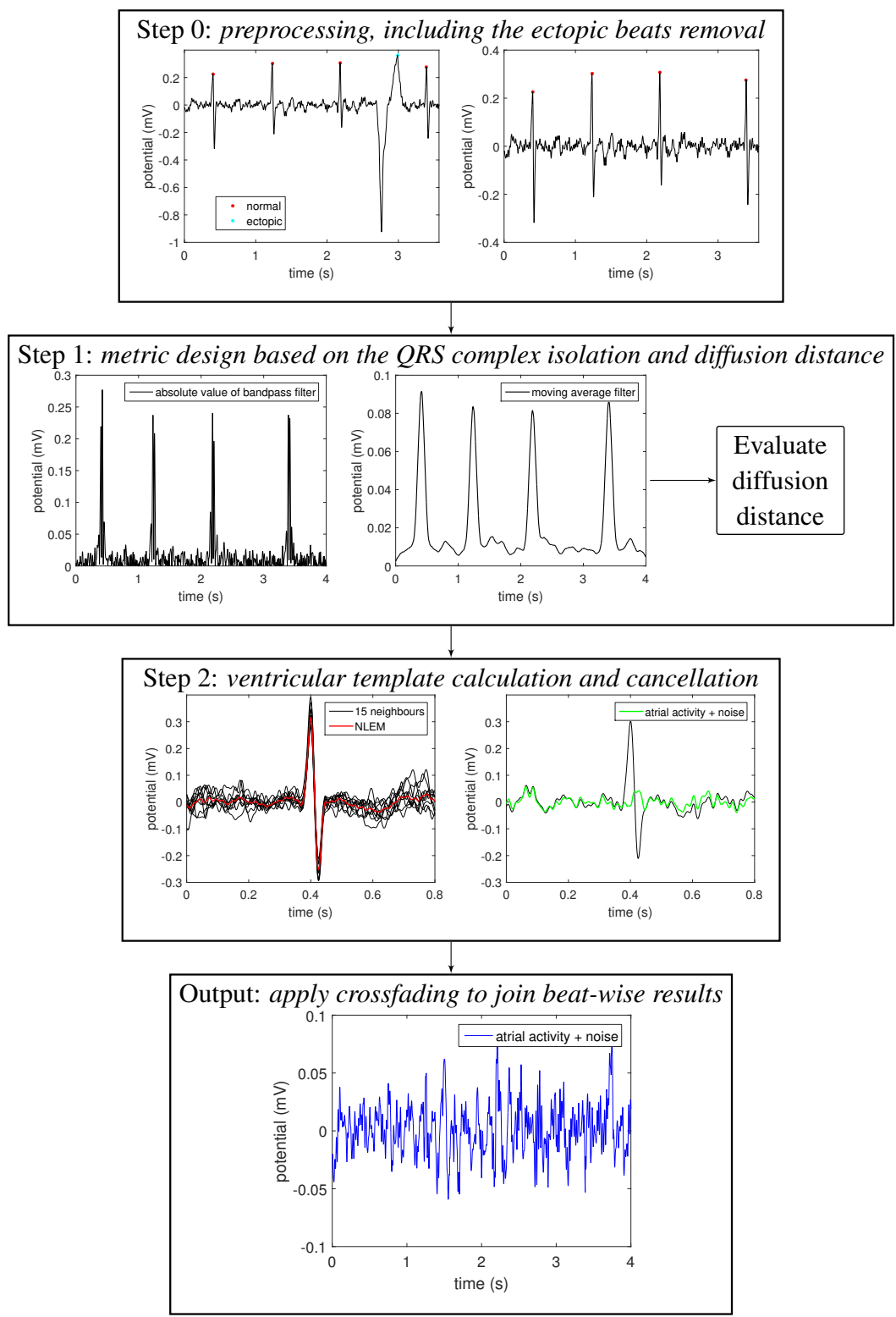
\begin{figure}
\centering
\begin{tikzpicture}[>=latex']
        \tikzset{block/.style= {draw, rectangle, align=center,minimum width=2cm,minimum height=.7cm,line width=0.3mm},
        lblock/.style={draw, rectangle, align=left,minimum width=2cm,minimum height=.7cm,line width=0.3mm},
        smallblock/.style={draw, rectangle, rounded corners = 1, align=center,minimum width=2cm,minimum height=.7cm,line width=0.18mm},
        input/.style={ 
        draw,
        trapezium,
        trapezium left angle=60,
        trapezium right angle=120,
        minimum width=2cm,
        align=center,
        minimum height=.7cm
    },
        }
        \node [block] (ectopic) {Step 0: \textit{preprocessing, including the ectopic beats removal} \\ \ \ \vspace{5pt} \includegraphics[width=.3\textwidth]{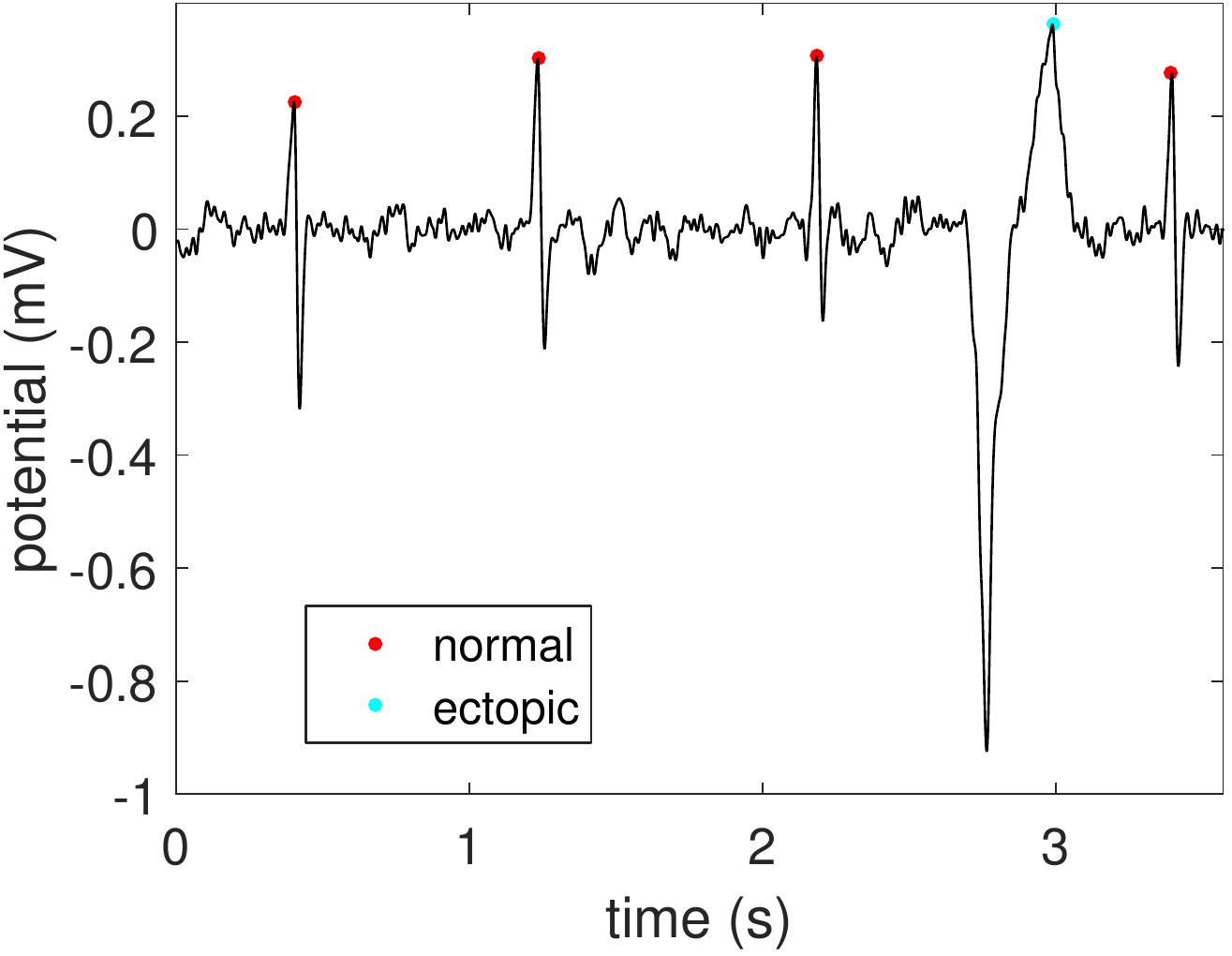} \ \ \includegraphics[width=.3\textwidth]{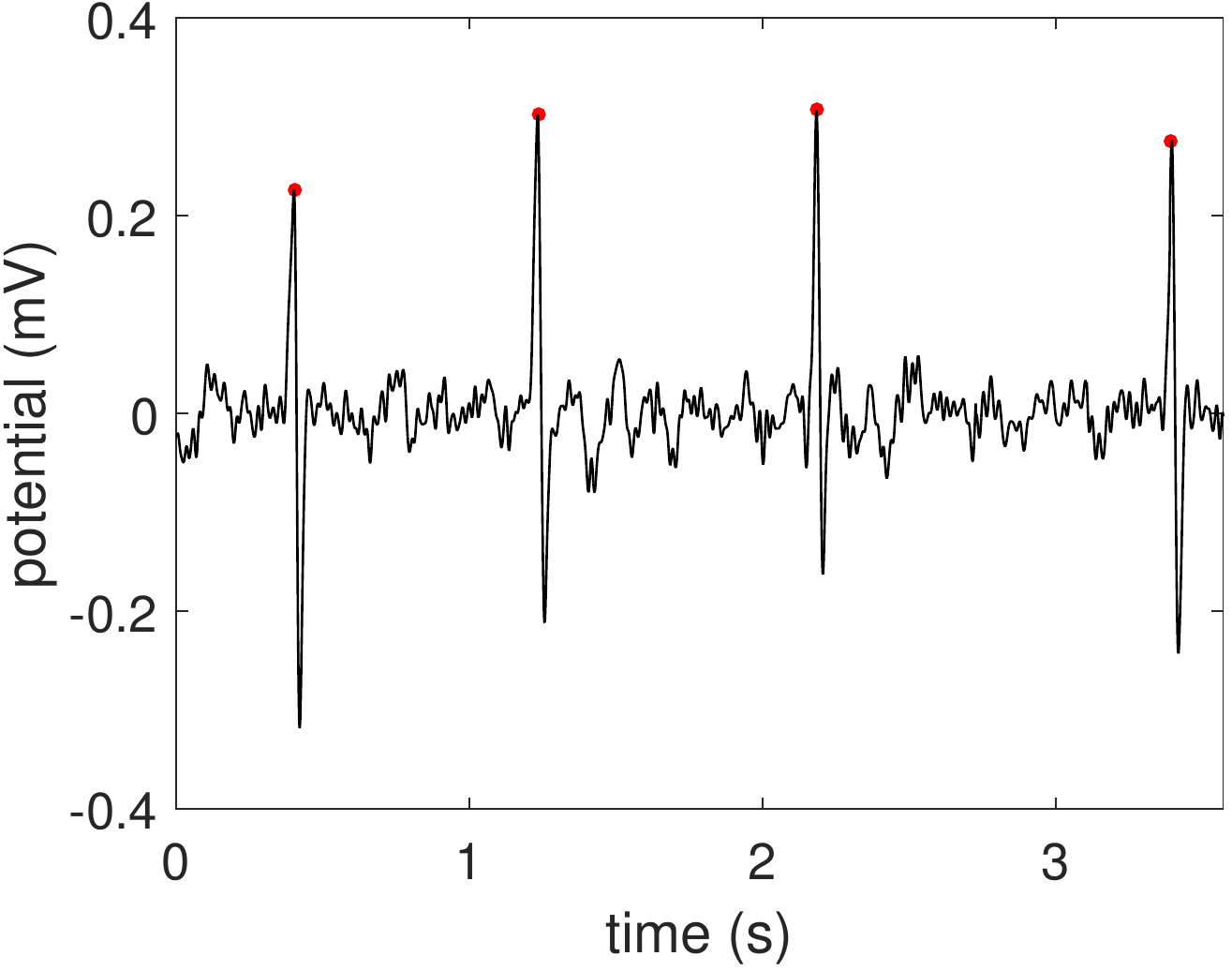} \ \ };   
           \node [lblock, below = 0.5 cm of ectopic] (QRS) {Step 1: \textit{metric design based on the QRS complex isolation and diffusion distance} \\ \ \ \vspace{5pt}\includegraphics[width=.3\textwidth]{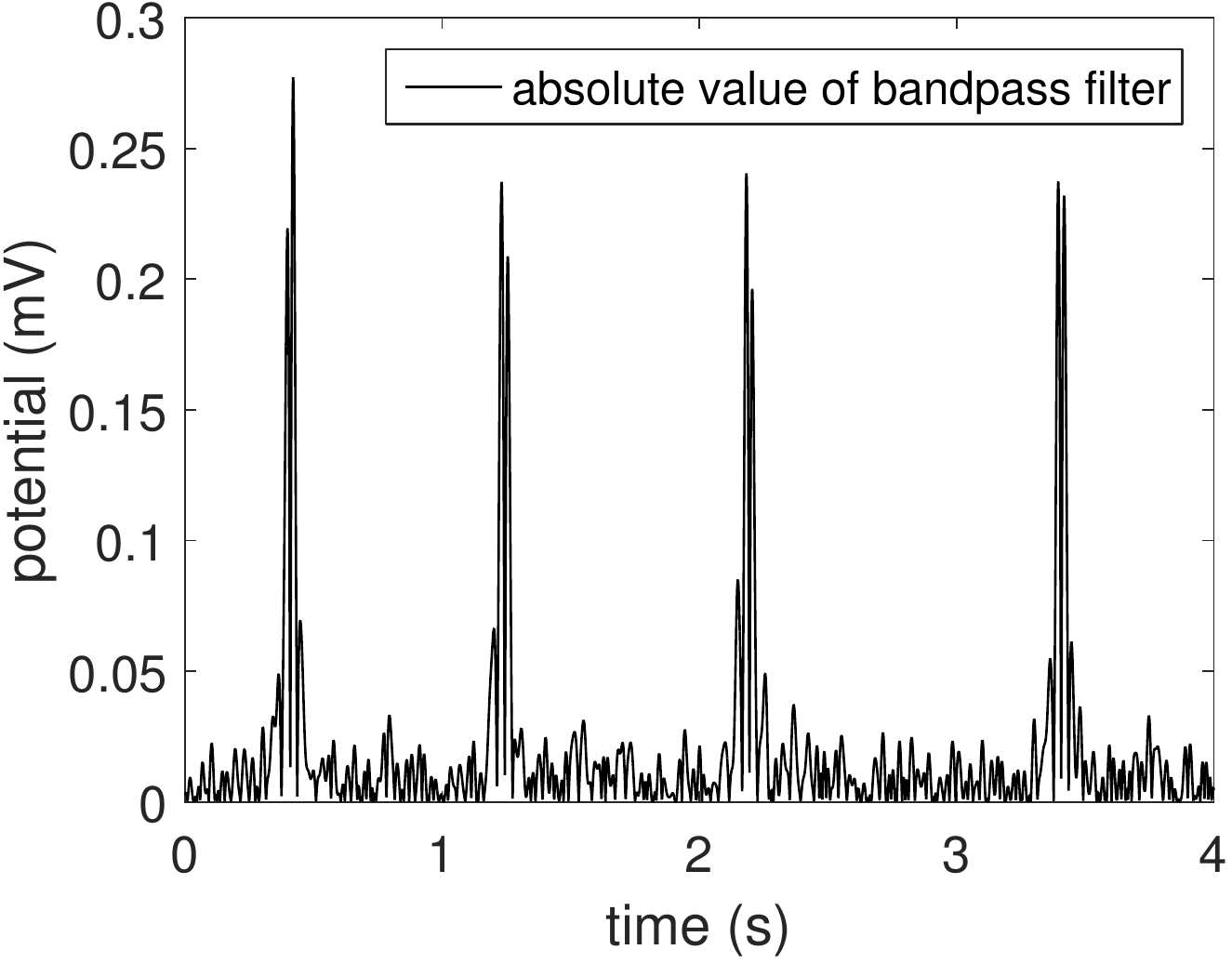} \ \ \includegraphics[width=.3\textwidth]{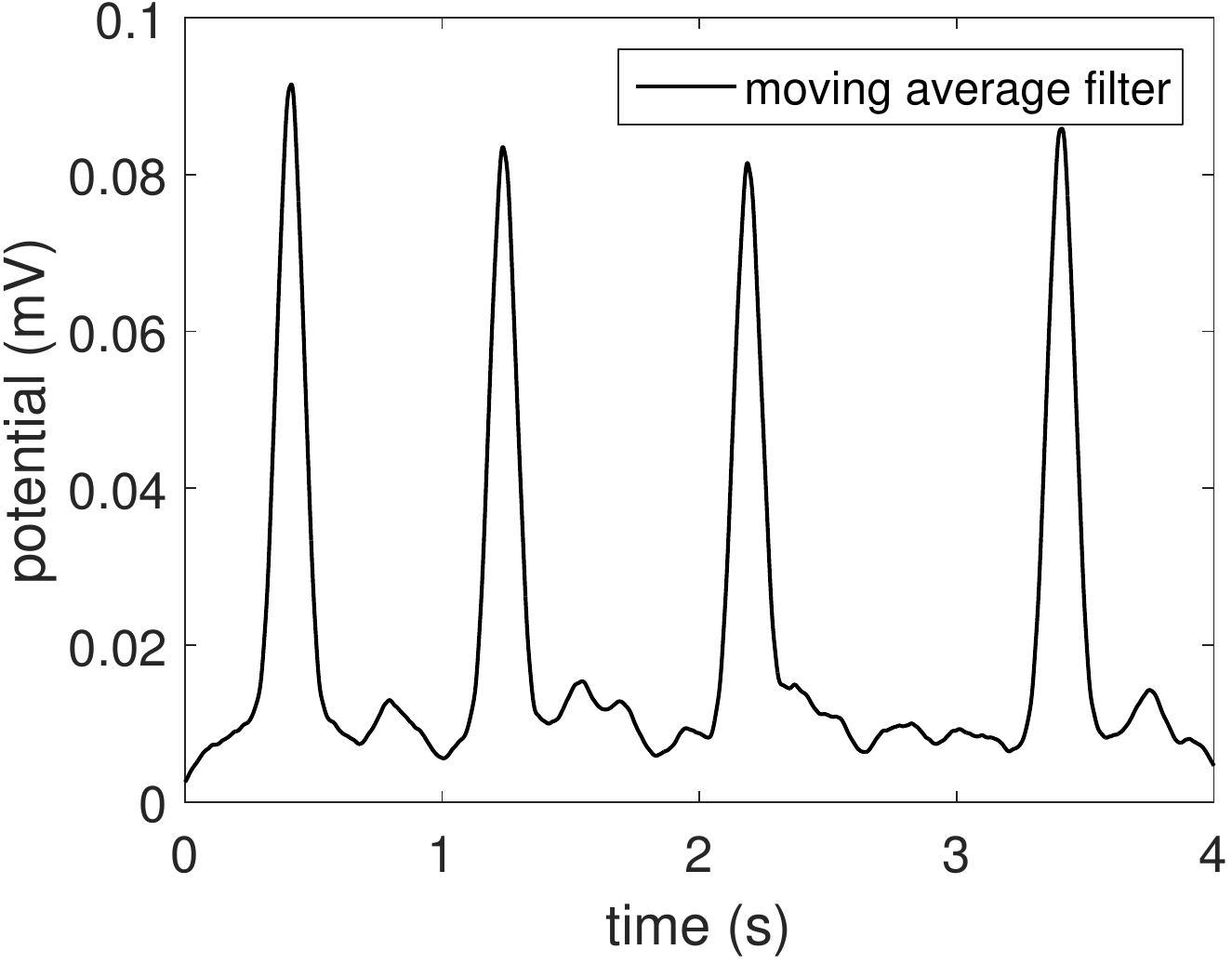}};
     \node [smallblock, right = -2.8 cm of QRS] (embed) {Evaluate \\ diffusion \\ distance };

        \node [block, below  =0.5 cm of QRS] (estimateventricular) {Step 2: \textit{ventricular template calculation and cancellation}\\ \ \ \vspace{5pt} \includegraphics[width=.3\textwidth]{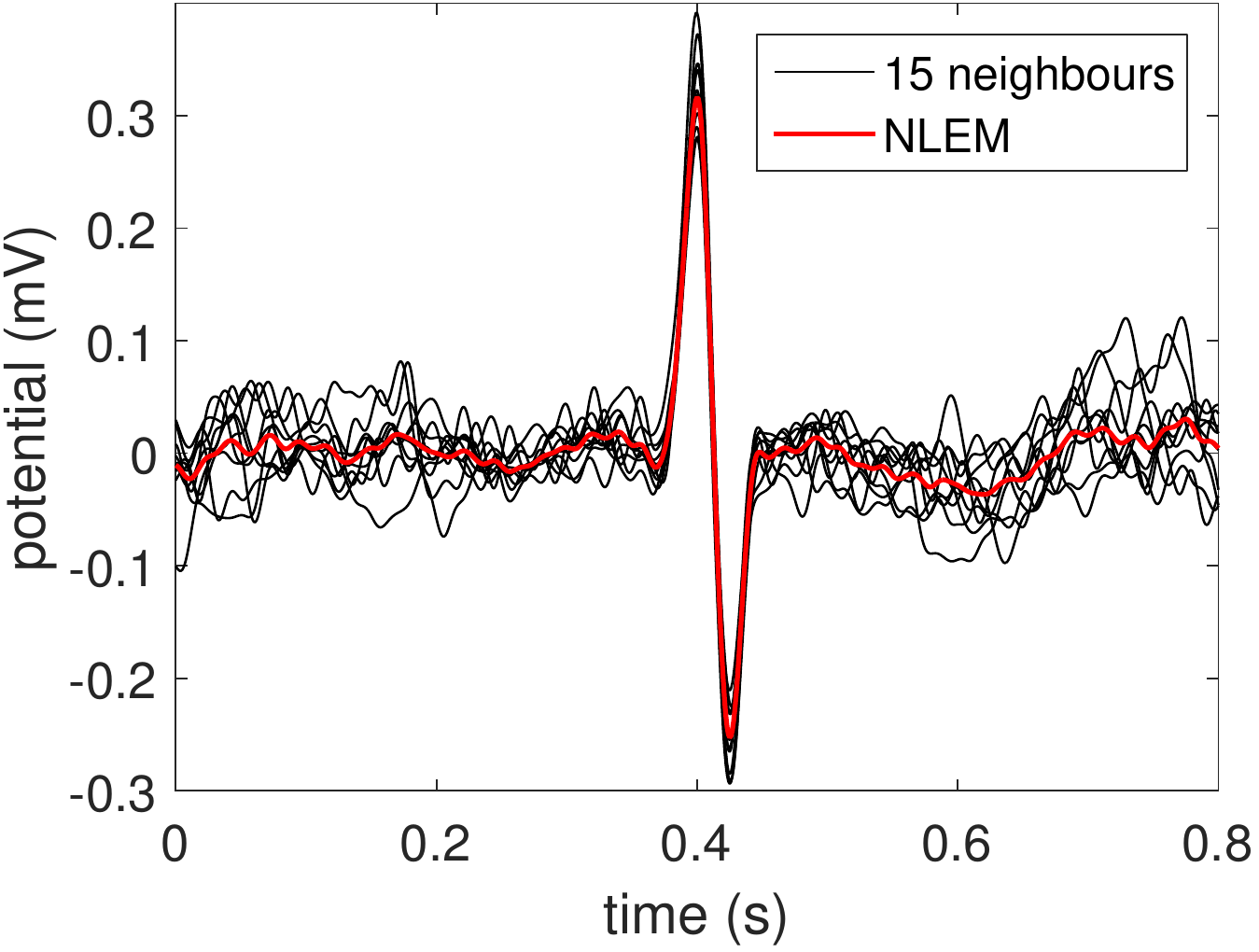} \ \  \includegraphics[width=.3\textwidth]{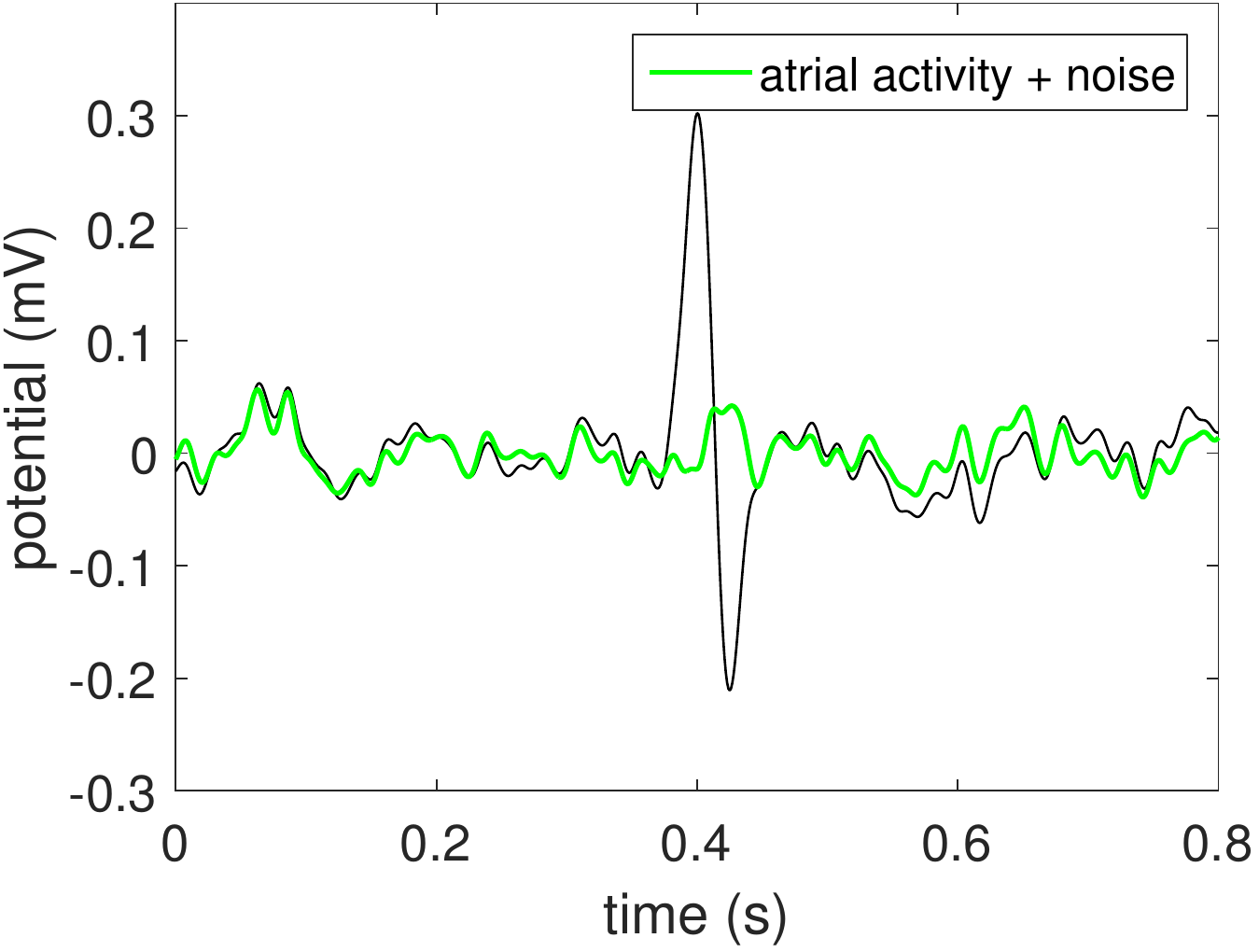}\ \  };
        
        \node [block, below =0.5 cm of estimateventricular] (end) { Output: \textit{apply crossfading to join beat-wise results} \\ \ \ \vspace{5pt} \includegraphics[width=.364\textwidth]{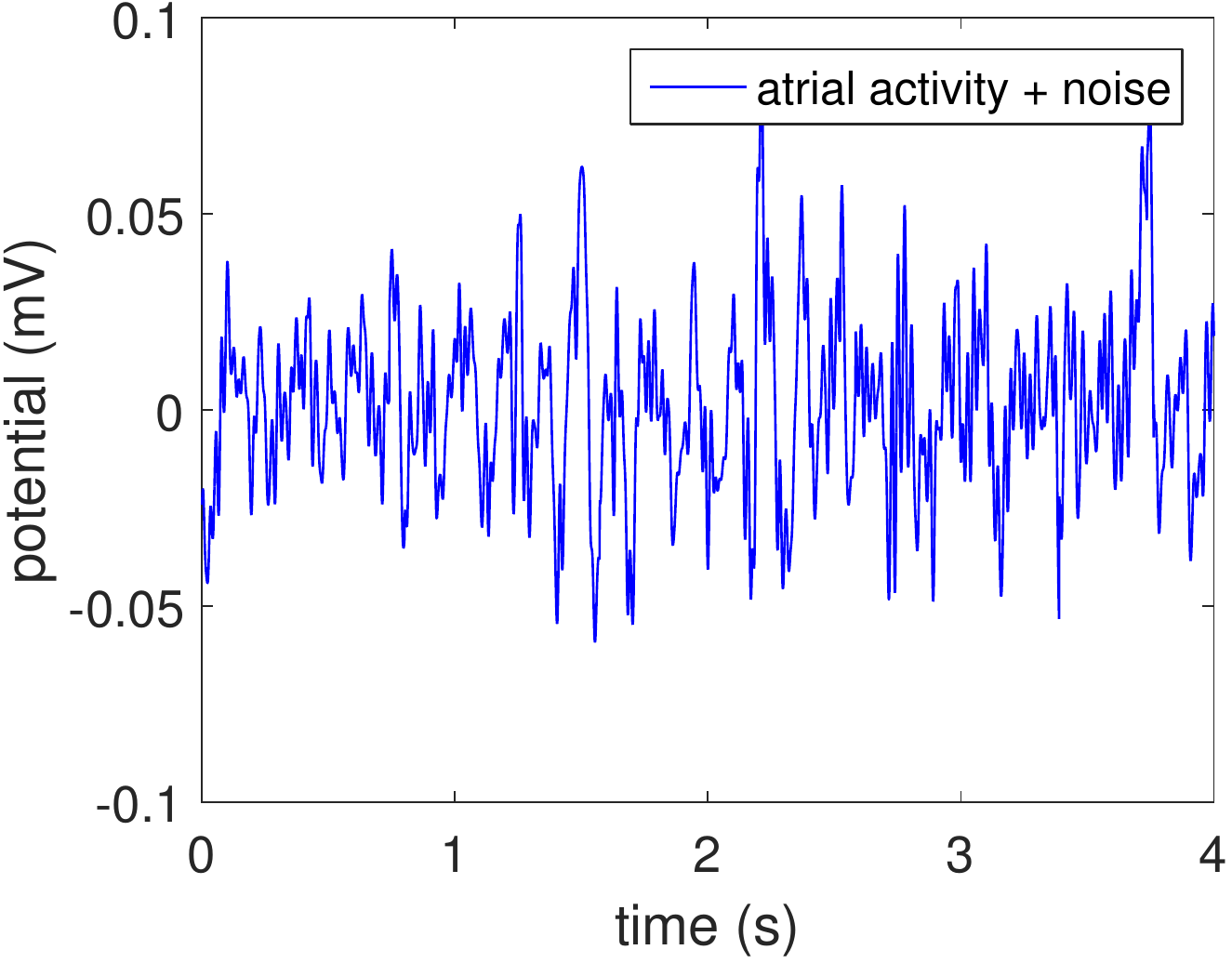}\ \  };

        \node [coordinate, above left = 0 cm and 2.3cm of ectopic] (topleft) {};
        \node [coordinate, above right = 0 cm and 1 cm of ectopic] (topright) {};
        \node [coordinate, right = 1 cm of ectopic] (rightembed) {};
        \node [coordinate, right = -3.84 cm of QRS] (DDarrow) {};
        
        \path[draw, ->] (DDarrow) -- (embed.west);
          \path[draw,->]   (estimateventricular.south) --(end.north);
        \path[draw,->] (ectopic.south)  --(QRS.north);
	\path[draw,->]   (QRS.south) --(estimateventricular.north);
       
\end{tikzpicture}
\caption{The step-by-step illustration of the proposed single-lead f-wave extraction algorithm, $\ours$.} \label{fig:summary1}
\end{figure}

There are essentially two steps in $\ours$, which could be understood as a generalization of the two essential steps in the ABS algorithm. 
The algorithm is summarized as follows. After pre-processing the ECG signal, we excise those regions containing ventricular activity. Equipped with a kernel, these excised segments constitute a point cloud in high-dimensional space. Our kernel attempts to identify segments that contain similar ventricular complexes. We integrate and denoise this kernel to define a metric called the \textit{diffusion distance} \cite{Coifman_Lafon:2006}, which characterizes the point cloud in terms of those morphological parameters which govern ventricular complex diversity most strongly. Finally, the ventricular activity in each segment is estimated by taking the Euclidean median of those segments which are its closest neighbours in terms of the diffusion distance.
An illustration of $\ours$ is shown in Figure \ref{fig:summary1}. 

\medskip
\textbf{Step 0: Pre-processing.}  The pre-processing step contains several commonly applied methods in the field.

\textit{Discretization.}
Let $x:\mathbb{R}\to \mathbb{R}$ denote the recorded single channel surface ECG signal, which satisfies 
\begin{equation}
x(t) = E(t) + \xi(t),
\end{equation}
where $E:\mathbb{R}\to \mathbb{R}$ is a continuous function describing the clean surface ECG signal and $\xi:\mathbb{R}\to \mathbb{R}$ is the noise encountered during the data collection process.  
We further decompose $E$ as 
\begin{equation}
E(t) = a(t) + v(t),
\end{equation}
where $a$ describes the recorded atrial activity and $v$ describes the recorded ventricular activity.
We model the {\em f-wave} by $a(t)$. 

The recorded surface ECG signal is sampled at a rate of $f_s$ samples per second, and is saved as a $N$-dimensional vector $\mathbf{x} \in \mathbb R^N$, where for each $i \in \{1, ..., N\}$, we have
\begin{equation}
\mathbf{x}(i) := x\left(\frac{i}{f_s}\right)\,,
\end{equation}
Clearly, the signal lasts for $T=n/f_s$ seconds.
The collected samples satisfy
\begin{equation}
\mathbf{x}=\mathbf{E}+\boldsymbol{\xi}=\mathbf{v} + \mathbf{a}+\boldsymbol{\xi}\,,
\end{equation}
where $\mathbf{E},\mathbf{v},\mathbf{a},\boldsymbol{\xi} \in \mathbb R^N$ are discretizations of $E$, $v$, $a$, and $\xi$ respectively. 
The goal of the proposed algorithm is to cancel $\mathbf{v}$ from $\mathbf{x}$ so that we obtain the discretized f-wave $\mathbf{a}$.

\medskip
\textit{Cleaning.}
We remove the baseline wandering using a two step process.  {We estimate the baseline wandering by first applying a median filter with window length 400 ms. A median filter is more useful than a moving average filter because it is less sensitive to high-amplitude peak information.  However, the output of a median filter typically contains unnatural jump discontinuities, so we subsequently apply a moving average filter with window length 400 ms to the output of the median filter to obtain a smoother baseline wandering estimator.} After removing the estimated baseline wandering from $\mathbf{x}$, the high frequency noise is suppressed using an order-60 bidirectional lowpass FIR filter with cutoff frequency 70 Hz. 
If needed, the power-line interference is suppressed using an order-60 bidirectional bandstop FIR filter at 60 Hz.
Contrasting the work in \cite{HamiltonCurley2000}, we use FIR filters to prepare for an implementation of the algorithm in real-time.
If $f_s$ is lower than $1000$ Hz, the signal is upsampled to $1000$ Hz to enhance R peak alignment \cite{Laguna2000,Bollmann2006}.
To preserve local information in the upsampling process and to prepare for an implementation of the algorithm in real-time, we apply the blending operator introduced in \cite{Chui_Lin_Wu:2016}, which is a local version of cubic spline interpolation.

\medskip
\textit{Annotation.}
We apply the R peak detection algorithm proposed by Elgendi \cite{Elgendi2013} to find all QRS complexes. Suppose there are $n$ R peaks, where $n\in\mathbb{N}$. We denote the index of the $i$th R peak by $t_i\in\mathbb{N}$.
An additional stage not in Elgendi's original paper but suggested in \cite{Herry_Frasch_Wu:2015} is included to reduce the incidence of false positives. When two peaks are detected within a refractory period of $250$ ms of one another, the peak with the sharpest slope is retained.  This modification is similar to a technique used in \cite{Pan_Tompkins:1985}. 
Finally, each QRS complex is classified as either ``normal'' or ``ectopic'' using the method proposed in \cite{Martinez2013}. 

\medskip
\textit{Ectopic beat removal.}  The methods recommended for ectopic beat removal by template subtraction are detailed here.
First, we choose $J_e \in \mathbb N$ so that a window of length $(2J_e + 1) / f_s$ seconds centered at each ectopic peak is large enough to envelope all of the ventricular activity associated with that peak.
Suppose there are $n_e$ ectopic beats, where $n_e\in\mathbb{N}$. We define the {\em ectopic patch} $\mathbf{e}_i \in \mathbb{R}^{2J_e + 1}$ associated with the $i$th ectopic beat to be 
\begin{equation} 
\label{ectopicregions}
\mathbf{e}_i = \big( \mathbf{x}\left( t_i - J_e \right),\, \dots ,\, \mathbf{x} \left( t_i \right),\, \dots ,\, \mathbf{x} \left( t_i + J_e \right) \big)^\top\in \mathbb{R}^{2J_e + 1}.
\end{equation}
For each $i=1,\ldots,n_e$, we calculate the $k_0$-nearest ectopic patches to $\mathbf{e}_i$ using the \textit{cosine affinity} $C$, defined as $C(\mathbf{e}_i, \mathbf{e}_j) := \frac{\mathbf{e}_i \cdot \mathbf{e}_j }{\Vert \mathbf{e}_i \Vert \Vert \mathbf{e}_j \Vert}$,
where $\cdot$ denotes the Euclidean inner product and $\Vert \, \cdot \, \Vert$ denotes the Euclidean norm.
This affinity measures the similarity of vectors while disregarding their ``size,'' and is suggested in \cite{Martinez2013}. 
Write $\{ \mathbf{e}_{i1}, ..., \mathbf{e}_{ik_0}\}$ for the $k_0$-nearest ectopic patches to $\mathbf{e}_i$.
The ventricular activity associated with the $i$th ectopic beat, denoted as $\tilde{\mathbf{v}}_i\in \mathbb{R}^{2J_e+1}$,  is estimated by taking the Euclidean median of the vectors $\{ \mathbf{e}_{i1}, ..., \mathbf{e}_{ik_0}\}$ by minimizing the following functional:
\begin{equation}
\tilde{\mathbf{v}}_i:=\argmin_{\mathbf{v}\in \mathbb{R}^{2J_e+1}}\sum_{j=1}^{k_0} \exp\left[ -\frac{ C(\mathbf{e}_i, \mathbf{e}_{ij})^2 }{h_0} \right] \|\mathbf{v}-\mathbf{e}_{ij}\|
\end{equation}
where $j = 1,\ldots, k_0$, and $h_0 > 0$ is a bandwidth parameter governing the influence of those points based on $C$. Denote by $\odot$ the Hadamard coordinate-wise product. We sequentially remove the ectopic beat templates from $\mathbf{x}$. 
{Define $\phi^L_c \in \mathbb R^{L + 1}$, where $c,L\in \mathbb{N}$ and $c<L/2$, to be a taper window defined as
\begin{equation}\label{Definition:taper:phi_0}
\phi^L_c(i) = \left\{  
\begin{array}{lll} \sin^2 \left[ \frac{\pi (i - 1)}{2c} \right]  & \textup{if} & 1 \leq i \leq c \ \\
1 & \textup{if} & c < i < L + 1 - c \ \\
\sin^2 \left[ \frac{\pi (i - L - 1)}{2 c} \right] & \textup{if} & L + 1 - c \leq i \leq L + 1,
\end{array} \right.
\end{equation}
and $c = \lfloor 0.1 f_s \rfloor\in \mathbb{N}$ is the chosen width for the tapering window. }
Starting from $i=1$, the VA is canceled by $\mathbf{e}_i - \phi^{2J_e}_{\lfloor 0.1 f_s \rfloor} \odot \tilde{\mathbf{v}}_i$.
%
%
We iteratively remove the VA in chronological order for $i=2,\ldots,n_e$.
Here, the taper window $\phi^{2J_e}_{\lfloor 0.1 f_s \rfloor}$ is chosen to avoid the boundary effect.
By removing the ventricular activities of the ectopic beats, we obtain the ectopic beat-free ECG signal. We use the same notation $\mathbf{x}$ and the R peak location index $t_i$, $i=1,\ldots,n$, to simplify the notation. 

\medskip
\textit{Excision of regions containing ventricular activity.}
Based on a priori physiological knowledge, we choose $I, J \in\NN$ so that the window over $[t_i-I,t_i+J]$ is long enough to include the ventricular activity associated with the $i$th R peak.
We define the {\em patch} $\mathbf{x}_i \in \mathbb{R}^{I + J + 1}$ associated with the $i$th ventricular activity to be 
\begin{equation} 
\label{regions}
\mathbf{x}_i = \big( \mathbf{x}\left( t_i - I \right),\, \dots ,\, \mathbf{x} \left( t_i \right),\, \dots ,\, \mathbf{x} \left( t_i + J \right) \big)^\top\in \mathbb{R}^{I+J + 1}
\end{equation}
so that the $i$th R peak is the $(I+1)$th coordinate of $\mathbf{x}_i$.
We view our collection of patches as a point cloud $\mathcal{X}=\{\mathbf{x}_i \}_{i=1}^n\subset\mathbb{R}^{I + J + 1}$, and we call the point cloud the {\em patch space}. In practice, we set $I$ small compared with $J$ because Af patients do not have P waves.  However $J$ must be large to capture the T wave.

\medskip
\textbf{Step 1: Metric design.} 
We intend to design a metric on $\mathcal{X}$ which depends only on the VA in each patch.  However, since each patch contains also the atrial wave and noise, any direct attempt based simply on the patch will be inaccurate. We now design a metric that depends only on the VA, and also effectively ignores the atrial wave while being robust to noise.

\medskip
\textit{Isolation of ventricular activity.} We perform an adaptation of the filtering performed in Elgendi's QRS detection algorithm to yield a signal which contains primarily QRS complex activity \cite{Elgendi2013}. We apply a bidirectional 3rd order 8-40 Hz bandpass Butterworth filter to $\mathbf{x}$, and set $\mathbf{s}=|\mathbf{x}| \in \mathbb R^n$, where $|\cdot|$ means taking the absolute value coordinate-wisely. We let $\mathbf{q}$ be the result of applying a moving average filter with window length 100 ms to the coordinate-wise square of $\mathbf{s}$. {In Figure~\ref{fig:summary1}, the box labeled ``Step 1'' depicts on the left the result of computing $\mathbf{s}$, and in the middle the result of computing $\mathbf{q}$.}
The purpose of taking the coordinate-wise square is to emphasize the high-frequency and high-amplitude content in $\mathbf{s}$.
Set $\mathbf{y}:=\mathbf{q}\odot\mathbf{s}$. We excise from $\mathbf{y}$ a collection of {\em surrogate patches} $\{\mathbf{y}_i \}_{i=1}^n$ defined as
\begin{equation}
\mathbf{y}_i = \big( \mathbf{y}\left( t_i - I \right),\, \dots ,\, \mathbf{y} \left( t_i \right),\, \dots ,\, \mathbf{y} \left( t_i + I \right) \big)^\top \in\mathbb{R}^{2I+1},
\end{equation}
where $I$ is as in (\ref{regions}).  Note that we do not consider $J$ since the T wave has been suppressed after multiplication by $\mathbf{q}$.

\medskip 
\textit{Diffusion distance.} Ideally, the surrogate patches $\mathbf{y}_i$ and $\mathbf{y}_j$ are similar if and only if the VA in $\mathbf{x}_i$ is similar to that in $\mathbf{x}_j$. 
However, it is not guaranteed since the VA information might be deformed in the surrogate patches $\{\mathbf{y}_i\}$, and the f-wave residua could not be totally removed. Thus, while the similarity between surrogate patches carries VA similarity information, it is contaminated by other inevitable factors. To handle this issue, we apply the DD, which is a metric robust to the ``noise.''  The DM and DD algorithms are described in detail in \ref{section:DM}. 

Fix $k_1\in\mathbb{N}$. For $\mathbf{x}_i$, find its $k_1$-nearest neighbours with the metric 
\begin{equation}
d_{\texttt{VS}}(\mathbf{x}_i,\mathbf{x}_j):=\| \mathbf{y}_i - \mathbf{y}_j \|,
\end{equation}
where we call $d_{\texttt{VS}}$ the {\em ventricular similarity metric}. 
Denote the set of the $k_1$-nearest neighbours of $\mathbf{x}_i$ as $\mathcal{N}_i$.
Define the sparse, symmetric affinity matrix $W\in\mathbb{R}^{n\times n}$ as 
\begin{equation}
W_{ij} = \Bigg\lbrace 
\begin{array}{ll} 
\exp\Big[-\frac{d^2_{\texttt{VS}}(\mathbf{x}_i, \mathbf{x}_j)}{h_1}\Big] & \textup{if } \mathbf{x}_i \in \mathcal{N}_j \mbox{ or } \mathbf{x}_j \in \mathcal{N}_i\\
0 & \textup{otherwise.}
\end{array}
\end{equation} 
where $h_1 > 0$ controls the influence of points which are further away in terms of the Euclidean distance. 
Define the diagonal degree matrix $D\in\mathbb{R}^{n\times n}$ as 
\begin{equation}
D_{ii} = \sum_{j \geq 1} {W_{ij}}.
\end{equation}
Define the transition matrix $P$ as
\begin{equation}
P = D^{-1} W,
\end{equation}
and write the right-eigenvectors of $P$ as $\varphi_1,\, ...,\, \varphi_n$, with their corresponding eigenvalues $\lambda_1\geq \cdots \geq \lambda_n$.
The DM with the diffusion time $t>0$ is given by 
\begin{equation}
\Phi_t: \mathbf{x}_i\mapsto \big(\lambda_2^t\varphi_2(i),\ldots,\lambda_{d+1}^t\varphi_{d+1}(i)\big)^\top\in\mathbb{R}^d,
\end{equation}
where $d\in\mathbb{N}$ is chosen by the user. With the DM, our final metric is defined as
\begin{equation}
d_{\texttt{DVS}}(\mathbf{x}_i, \mathbf{x}_j) := \Vert \Phi_t(\mathbf{x}_i) - \Phi_t(\mathbf{x}_j) \Vert,
\end{equation}
which we call the {\em diffusion ventricular similarity metric}.
The resulting diffusion ventricular similarity metric will be used to organize $\mathcal{X}$, since it depends mainly on the VA, and is robust to the existing f-wave and noise. 

\medskip
\textbf{Step 2: Template calculation and removal.} 
With the designed metric $d_{\texttt{DVS}}$, we finally are able to evaluate the VA in each $\mathbf{x}_i$, and recover the f-wave. We propose to apply the non-local Euclidean median (NLEM) to achieve the goal.  {The NLEM is computed using an iteratively reweighted least squares-type algorithm as described in \cite{Chaudhury_Singer:2012}.}

\medskip
\textit{Non-local Euclidean medians.} 
Choose a number of neighbours $k_2\in \mathbb{N}$.
For each patch $\mathbf{x}_i$, choose the $k_2$-nearest patches, denoted as $\{\mathbf{x}_{i1},...,\mathbf{x}_{ik_2} \} \subset \mathcal{X}$ according to $d_{\texttt{DVS}}$. 
Note that $\mathbf{x}_{i1} = \mathbf{x}_i$. 
Denote $\tilde{\mathbf{v}}_i$ to be the estimated VA corresponding to $\mathbf{x}_i$ via evaluating the Euclidean median of the patches $\{\mathbf{x}_{i1},...,\mathbf{x}_{i k_2} \}$:
\begin{equation}\label{Algorithm:FinalStep:NLEM}
\tilde{\mathbf{v}}_i = \argmin_{\mathbf{v}\in \mathbb{R}^{I+J+1}} \sum_{j = 1}^{k_2} \exp\left[ -\frac{ d_{\texttt{DVS}}(\mathbf{x}_i, \mathbf{x}_{ij})^2 }{h_2} \right] \Vert \mathbf{x}_{ij} - \mathbf{v} \Vert\,,
\end{equation}
where $h_2 > 0$ is a bandwidth parameter chosen by the user.
The f-wave activity in the $i$th patch is then estimated to be
\begin{equation}
 \tilde{\mathbf{a}}_i =\mathbf{x}_i-\tilde{\mathbf{v}}_i.
\end{equation}

\textit{Crossfading.}  We join the beat-wise results $\{\tilde{\mathbf{a}}_i\}_{i=1}^n$ to estimate the f-wave activity in $\mathbf{x}$. Initially, set $\tilde{\mathbf{a}} = \mathbf{x}$, and then subsequently substitute
\begin{equation}
\tilde{\mathbf{a}}(t_i - I, \, ... , \, t_i + J) ={\phi^{I+J}_{\lfloor 0.1 f_s \rfloor} } \odot \, \tilde{\mathbf{a}}_i + (1 - {\phi^{I+J}_{\lfloor 0.1 f_s \rfloor}}) \odot \, \mathbf{x}_i,
\end{equation}
for every $i=1,\ldots,n$.
In other words, we obtain a blending of the estimate f-wave with the patch $\mathbf{x}_i$ near the edges of $\mathbf{x}_i$.  

\begin{remark}
Note that we frequently observe uncanceled QRS complexes at the end of the patches $\mathbf{a}_i$ in situations where the heart beat is rapid {because the original patches $\mathbf{x}_i$ may overlap}. Thus substitution of the patches $\{ \mathbf{a}_i \}$ should be done iteratively in chronological order.
\end{remark}

\subsection{A variation of $\ours$}\label{subalgorithm2}

In some cases, we may only have a short ECG recording. In this case, running DM might not be feasible. In this situation, we could replace (\ref{Algorithm:FinalStep:NLEM}) by
\begin{equation}\label{Algorithm:FinalStep:NLEM}
\tilde{\mathbf{v}}_i = \argmin_{\mathbf{v}\in \mathbb{R}^{I+J+1}} \sum_{j = 1}^{k_2} \exp\left[ -\frac{ d_{\texttt{VS}}(\mathbf{x}_i, \mathbf{x}_{ij})^2 }{h_2} \right] \Vert \mathbf{x}_{ij} - \mathbf{v} \Vert\,,
\end{equation}
that is, we use the $d_{\texttt{VS}}$ metric to determine the neighbors. We call such variation $\oursX$.

\begin{remark}
For the theoretical justification supporting the usefulness of DD, we refer the readers with interest to \ref{Section:Model}. We remark that a branch of mathematics called differential geometry provides the mathematical support to design this metric. Mathematically, $\mathcal{X}$ could be modeled as a set coming from sampling a fiber bundle. The {\em base manifold} is parameterized according to ventricular complex morphology, which is constrained by the physiological ``homeostasis''. Under a mild assumption, the f-wave contamination is modeled as the fiber attached to the base manifold. The fiber bundle structure we could describe based on our knowledge is trivial; that is, the fiber bundle is a direct product of the base manifold and the fiber. The metric $d_{\texttt{VS}}$ and $d_{\texttt{DVS}}$ designed above is attempting to ignore the fiber and obtain an estimate of the distance intrinsic to the manifold.
\end{remark}

\subsection{Material}\label{Section:material}

To validate the proposed algorithm, we consider two different types of simulated signals and a set of real Holter signals. 
We prepare simulated signals by adding an atrial wave and noise to a purely ventricular signal.

\subsubsection{First simulated f-wave}\label{Section:Simulation1}

The first method involves excising the ventricular segments of a real ECG signal and splicing together the remnants, as suggested in \cite{Alcaraz2008}. We provide the details here. Obtain a real ECG signal from an Af patient sampled at 1000 Hz, remove baseline wandering using the combined median and moving average filters used described in the pre-processing stage of our algorithm, and obtain the RR intervals by cutting the signal at each R peak. Remove any RR intervals containing ectopic beats. Remove any RR intervals not exceeding 800 ms in length. Remove the first 400 ms and the last 200 ms from each RR interval to excise the ventricular activity.  Concatenate the obtained segments in chronological order. To deal with sudden changes in the atrial wave's amplitude at points where concatenation took place, apply a 6th-order lowpass Butterworth filter with cutoff frequency 15 Hz to the entire signal.  

The noise of the simulated ECG is created by applying a 12-70 Hz bandpass filter to Gaussian white noise and scaling the standard deviation of the filtered noise to be 0.003 mV plus $5\%$ the standard deviation of the atrial wave.

\subsubsection{Second simulated f-wave}\label{Section:Simulation2}

The second f-wave simulation is generated using a generalization of the sawtooth model \cite{Stridh2001,Petrenas2012}.
To obtain a more realistic f-wave, we generalize the sawtooth model by taking the adaptive non-harmonic model \cite{lin2016waveshape} into account. 
Choose an initial amplitude $a_0>0$ for the atrial wave, and modulate this amplitude by composing a sine wave having amplitude $\Delta_a>0$ with a Gaussian random walk $\varphi \colon \mathbb N \rightarrow \mathbb R$ whose steps are sampled from a Gaussian distribution with mean zero and standard deviation $\alpha>0$. For $t\in \mathbb N$, we write
\begin{equation}
a_1(t) = a_0 + \Delta_a \sin(\varphi(t)).
\end{equation} 
Similarly, choose a fundamental frequency $\xi_0>0$ for the atrial wave, and modulate this frequency by choosing an amplitude $\Delta_\xi>0$ and subsequently writing
\begin{equation}
\xi_1(t) = \xi_0 + \Delta_\xi \sin (\psi(t)),
\end{equation} 
where $t \in \mathbb N$ and $\psi \colon \mathbb N \rightarrow \mathbb R$ is a Gaussian random walk whose steps are sampled from a Gaussian distribution with mean zero and standard deviation $\beta>0$.  Choose a number $M$ of multiples to simulate the ``sawtooth'' shape. The frequency of the $k$th harmonic, where $k=1\, ,\ldots,\, M$, is the $k$th multiple of $f_0$.  These harmonics are added with successively smaller amplitudes, and their amplitudes and frequencies are varied proportionally. Indeed, we set 
\begin{equation}\label{simulation}
a(t)=\sum_{m=1}^M \frac{1}{m}a_1(t) \sin\left(2 \pi m\int_0^t \xi_1(s) \, ds \right),
\end{equation}  
where $t \in \mathbb N$, which is discretized at the sampling rate $f_s$ by  
\begin{equation}
\mathbf{a}(i)=\sum_{m=1}^M \frac{1}{m}a_1(t/f_s) \sin\left(\frac{2 \pi m}{f_s} \sum_{j = 0}^i \xi_1(j/f_s) \right).
\end{equation}
To combat the regularity of the results, we apply a  2-7 Hz bandpass filter to Gaussian white noise, scale the standard deviation of the filtered noise to be $\zeta$\% of the standard deviation of the atrial wave, and add it to the signal. We invert the signal over the horizontal axis with probability 0.5. We apply a 6th-order lowpass Butterworth filter at 15 Hz to smooth the result. Note that the average amplitude of the signal will in general be larger than $a_0$ due to the addition of superpositioned overtones and noise.  
 
The noise of the simulated ECG is created by applying a 12-70 Hz bandpass filter to Gaussian white noise and scaling the standard deviation of the filtered noise to be 0.003 mV plus $5\%$ the standard deviation of the atrial wave. 

{The generalization of the model in \cite{Stridh2001,Petrenas2012} lies in the randomness of the modulation. Note that in our model, the amplitudes and frequencies are varied according to a random process, which results in less regularity in the signal. Furthermore, all oscillations generated in the sawtooth model are not symmetric, and inverting over the horizontal axis yields a different signal. By taking this into account, the resulting signal will be less regular and further catch the complexity of the f-wave. The above consideration is especially relevant when simulating long f -wave signals.}

\subsubsection{Simulated ventricular activity}\label{Section:SimulationV}

Finally, we obtain ventricular activity simulations using the dynamical model ECGSYN developed in \cite{McSharry2003}. We generate signals lacking P waves, and we vary the amplitudes, positions, and widths of the Q, R, S, and T waves from signal to signal. The implementation of ECGSYN that we use performs QT interval stretching according to Bazett's formula. In practice, the relationship between RR interval length and QRST complex amplitude is noisy. The ventricular signals are thus integrated beat-wise, and the amplitude of each QRST complex is scaled by a factor of $1 \pm 0.05$. (In some cases, T wave amplitude is non-linearly related to QRS complex amplitude, but we do not simulate this relationship.)  To simulate small perturbations in the cardiac axis from beat to beat, we modify the $z$-positions of the Q, R, and S peaks in each beat by adding $0 \pm 2$. Note that our synthetic ventricular signals do not contain ectopic beats, and that after adding the f-wave, we remove baseline wandering using the combined median and moving average filters described in the pre-processing stage of our algorithm. We assemble the final ECG signals by combining the simulated f-waves with these simulated ventricular signals.

\subsubsection{Real Holter signals} 

To evaluate the potential of applying the proposed algorithm to real data, we take one hour continuous ECG signals from 47 patients with persistent Af visiting Chang Gung Memorial Hospital during 2011. Patients received 24-hour Holter ECG recordings for evaluating cardiac arrhythmia, rate control of atrial fibrillation, or surveying symptoms related to rhythm disturbances.
This is a retrospective study, and Chang Gung Medical Foundation Institutional Review Board approved the study protocol (number: 103-7449B and 104-7769C). The data is collected from DigiTrak XT (Philips), and we focus on the channel 1, E(+) to S(-), under the EASI\textsuperscript{\textregistered} lead system.

\subsection{Other algorithms}\label{Section:Other}

Our proposed algorithm seeks to improve upon the performance of several existing algorithms, including ABS \cite{Slocum1985,Slocum1992,Holm1998,Shah2004}, local principal component analysis (PCA) \cite{Castells2005}, local adaptive singular value cancellation (aSVC)  \cite{Alcaraz2008} and non-local aSVC \cite[Section 3.3.]{Alcaraz2008}.  We describe these algorithms in detail below.

In each of these algorithms, we perform step 0 (signal pre-processing) as described in Section \ref{subalgorithm}.  These algorithms differ only when it comes to ventricular template selection.  The ABS, local PCA, and local aSVC algorithms obtain temporally local estimates for the VA in each beat.
In ABS, the ventricular template  
is the average of the first $k'\in\mathbb{N}$ patches to either side of $\mathbf{x}_i$, where $k'$ is chosen by the user. 
In local PCA and local aSVC, for each patch $\mathbf{x}_i$, we evaluate the singular value decomposition of the data matrix $\mathbf{X}$,  where 
\begin{equation}
\mathbf{X} = \begin{bmatrix} \mathbf{x}_{i - k'} & \cdots & \mathbf{x}_{i + k'} \end{bmatrix}\in \mathbb{R}^{(I+J+1)\times (2k'+1)}.
\end{equation}
We then set the template to be the top left singular vector (which is related to the first principal component of $\mathbf{X}$).
The main difference between local PCA and local aSVC is that in aSVC the amplitude distances between the Q and R points is scaled adaptively when canceling the VA \cite[(4)]{Alcaraz2008}. 

In non-local PCA, the cosine affinity is used to determine the neighbours which will be used for template calculation. Define $\{ \mathbf{x}_{i1}, ..., \mathbf{x}_{i(2k'+1)}\}$ to be the $(2k'+1)$-nearest patches to $\mathbf{x}_i$ in terms of the cosine affinity. Then,
estimate the VA in the patch $\mathbf{x}_i$ by the top left singular vector of the data matrix
\begin{equation}
\mathbf{X}:=\begin{bmatrix}
\mathbf{x}_{i1} &\cdots & \mathbf{x}_{i(2k'+1)}
\end{bmatrix}\in \mathbb{R}^{(I+J+1)\times (2k'+1)}.
\end{equation}
The approach in non-local aSVC is similar, where the neighbours are calculated using the correlation coefficient instead of the cosine distance, and the templates are adapted beat-wise based on the QR height.  Once the estimated templates have been obtained, we remove them from the patches $\{ \mathbf{x}_i \}$ and do crossfading as described in step 2 of our algorithm.  A graphical overview of the algorithms we will consider is provided in Table~\ref{methodcomparison}.

\begin{table}
\small
\begin{tabular}{|c|c|c|c|c|c|c|}
\hline
 & ABS \cite{Slocum1985} & local PCA \cite{Castells2005} &   non-local PCA & $\oursX$ & $\ours$ \\
 & &  and aSVC \cite{Alcaraz2008}& and aSVC \cite{Alcaraz2008} & &\\
 \hline\hline
 Neighbours & local  & local & non-local & non-local & non-local \\
 \hline 
Metric & N/A &  N/A & cosine affinity & ventricular  & diffusion ventricular \\ 
design & & &  & similarity metric & similarity metric \\
 \hline
 Template  & mean & principal    & principal  & Euclidean  & Euclidean  \\
 design & & components & components & median & median\\
 \hline
\end{tabular}
\caption{Comparison of various algorithms from three aspects. ABS: averaged beat cancellation/subtraction; PCA: principal component analysis; aSVC: adaptive singular value cancellation; NLEM: non-local Euclidean median; N/A: not applicable.}\label{methodcomparison}
\end{table}

\subsection{Evaluation metrics for simulated dataset}\label{Section:metrics} 

We will use multiple metrics to quantify the accuracy of ventricular cancellation in the simulated case.
The normalized mean square error (NMSE) \cite[Equation (16)]{CastellsRieta2005} between the simulated f-wave $\mathbf{a}\in\mathbb{R}^N$ and the extracted f-wave $\tilde{\mathbf{a}}\in\mathbb{R}^N$ is defined as 
\begin{equation}
\textup{NMSE} = { \frac{\sum_{i = 1}^N ( \mathbf{a} (i ) - \tilde{\mathbf{a}} ( i ) )^2}{\sum_{i = 1}^N \mathbf{a}(i) ^2}}.
\end{equation}
We also consider the cross-correlation of $\mathbf{a}$ and $\tilde{\mathbf{a}}$, denoted as $\rho$ \cite[Equation (16)]{Alcaraz2008}, to evaluate the reconstructed f-wave. 
We further evaluate the signal-to-noise ratio (SNR) and the peak signal-to-noise ratio (PSNR) commonly considered in the literature using the formulas 
\begin{equation}
\textup{SNR} = 20 \log_{10} \left[\frac{\texttt{std}(\mathbf{a})}{\texttt{rmse}(\mathbf{a},\tilde{\mathbf{a}})} \right] ;\ \ \  \textup{PSNR} = 20 \log_{10} \left[\frac{ \max_i| \mathbf{a} (i)|} {\texttt{rmse}(\mathbf{a}, \tilde{\mathbf{a}})} \right],
\end{equation}
where $\texttt{rmse}$ is the root mean square error.  Note that if $\mathbf{a}$ and $\mathbf{\tilde{a}}$ have the same mean, then $\texttt{rmse}(\mathbf{a}, \tilde{\mathbf{a}}) = \texttt{std}(\mathbf{a}- \tilde{\mathbf{a}})$.

\subsection{Evaluation metrics for real dataset}\label{Section:metrics2} 

In the real case, where the true f-wave is not available, we first consider the ventricular residue (VR) index proposed in \cite[Equation (18)]{Alcaraz2008}, which intends to evaluate how well the QRST complex is canceled. 
The VR index for the $i$th ventricular complex is defined as 
\begin{equation}\label{Definition:VRindex}
\textup{VR}_i = \frac{n}{ \sum_{l = 1}^n \tilde{\mathbf{a}}(l)^2 } \sqrt{\sum_{l = t_i - H}^{t_i + H} \tilde{\mathbf{a}}(l)^2} \max_{t_i - H \leq l \leq t_i + H} \vert \tilde{\mathbf{a}}(l) \vert,
\end{equation}
where $H = \lfloor 0.05 f_s \rfloor$. When the ventricular activity is not well removed and some big error exists, the VR index is large.
While the VR index provides information about the ventricular residue, however, it has several limitations. The most problematic one is its bias to the ``over-smoothing.'' If we remove all cardiac activity over the ventricular activity interval, then the VR index is $0$. In short, while the VR index could catch the big ventricular activity cancellation error, it might be problematic when the f-wave is wrongly extracted. We thus need an index that can simultaneously catch the ventricular and atrial activities correctly.
Inspired by the VR index and the above-mentioned needs, we propose the following {\em modified VR (mVR)} index. For the $i$th cardiac activity, define $\textup{mVR}_i$ as 
\begin{align}
\textup{mVR}_i:=&\frac{1}{2} \left[ \frac{ \texttt{med}|\tilde{\mathbf{a}}_i - \texttt{med}(\tilde{\mathbf{a}}_i)|}{\texttt{med}|\mathbf{a}_i - \texttt{med}(\mathbf{a}_i)|}+\frac{\texttt{med}|\mathbf{a}_i - \texttt{med}(\mathbf{a}_i)|}{\texttt{med}|\tilde{\mathbf{a}}_i - \texttt{med}(\tilde{\mathbf{a}}_i)|} \right]\nonumber\\
 &\times\frac{1}{2} \left[ \frac{\texttt{q95}|\tilde{\mathbf{a}}_i - \texttt{med}(\tilde{\mathbf{a}}_i)|}{\texttt{max}|\mathbf{a}_i - \texttt{med}(\mathbf{a}_i)|}+\frac{\texttt{max}|\mathbf{a}_i - \texttt{med}(\mathbf{a}_i)|}{\texttt{q95}|\tilde{\mathbf{a}}_i - \texttt{med}(\tilde{\mathbf{a}}_i) |} \right],
\end{align}
where $\texttt{med}$ is the median operator, $\texttt{q95}\mathbf{w}$ is the $95$th quantile of the absolute value of the vector $\mathbf{w}$, $\tilde{\mathbf{a}}_i$ is the estimated f-wave over the period of the $i$th cardiac activity, and $\mathbf{a}_i$ is the concatenation of true f-wave over the interval without any ventricular activity associated with the $i-30,\ldots,i+30$ beats. 
Note that the first term measures how the ``spectrum'' of the atrial activity is overall recovered by the algorithm. This term is introduced to prevent ``over-smoothing.'' 
The second term measures the presence of ventricular residua. This index captures simultaneously how well the ventricular activity is removed and how well the atrial activity is preserved. 

We also consider the {\em spectral concentration} (SC) index proposed in \cite[Equation (16)]{CastellsRieta2005}. The one-sided power spectral density $P \in \Bbb R^{4001}$ of $\tilde{\mathbf{a}}$ is calculated using Welch's method, featuring 8000 discrete Fourier transform points, Hamming windows of 400 samples, and 50\% overlapping.  The SC index for a signal sampled at 1000 Hz is defined as 
\begin{equation}
\text{SC} = \frac{\sum_{i = 24}^{96} P(i)}{\sum_{i=1}^{4001} P(i)}
\end{equation}
and is the percentage of energy in the frequency range $3$-$12$ Hz. 
This band is chosen to be sufficiently wide to cover the f-wave spectrum.

\subsection{Parameters}\label{paramopt}

For the purpose of reproducibility, we summarize the parameters used to generate our results. We make \textit{ad hoc} choices because the prospect of performing an optimization procedure such as a grid search would likely cause overfitting to either the database, the chosen evaluation metric (if we use real signals), or the simulation method (if we use simulated signals).  
These parameters are also fixed unless otherwise stated.  

The ECG signal is resampled to have the sample rate $f_s=1000$ Hz, and the data set is of length $T=3600$ seconds.
The parameters for the patch size are set to $J_e = 0.5$, $I=0.3$, and $J=0.8$ seconds. In Step 0, the
number of ectopic neighbours in the ectopic beat cancellation is set to $k_0=30$ and the ectopic bandwidth $h_0$ is defined point-wise for each $\mathbf{e}_i$ as $2C(\mathbf{e}_i, \mathbf{e}_{i8})^2$.
In Step 1, the number of neighbours is set to $k_1=15$ and the bandwidth $h_1$ is set as twice the square of the median $\ell^2$ distance to the $500$th neighbour. We take $N=2000$ neighbours to evaluate the affinity matrix and the embedding dimension is set to $d=30$.
In Step 2, the number of neighbours is set to $k_2=15$, the bandwidth $h_2$ is defined point-wise for each $\mathbf{x}_i$ as $2d_{\texttt{DVS}}(\mathbf{x}_i, \mathbf{x}_{i4})^2$, {the diffusion time $t$ is chosen to be $1$, and the embedding dimension $d$ is chosen to be $30$}.

The algorithms ABS, local PCA, non-local PCA, local aSVC, and non-local aSVC require fixing the number of neighbours. This parameter is fixed at $k'=7$ so that $2k' + 1 = 15$ unless otherwise stated.

The parameters for the sawtooth f-wave model discussed in Section \ref{Section:material} are summarized here.
The sample rate is $f_s=1000$ Hz, the number of samples is $n=3.6 \times 10^6$, the 
number of harmonics is $M=3$, the initial amplitude is set to $a_0=0.06 \pm 0.02$ mV, the change in amplitude is $\Delta_a=a_0 / 2$, the step size 1 is $\alpha=0.02$, the initial frequency is $f_0=6\pm 2$ Hz, the change in frequency is $\Delta_f=f_0 / 3$, the step size 2 is $\beta=0.1$, and the $2$-$7$ Hz noise is $\zeta=50\pm 20$\%.

The parameters for the ventricular simulations in ECGSYN are summarized here.
 The measurement noise is set to 0, the mean heart rate is set to $70 \pm 10$ bpm with the standard deviation $10$ bpm, the low frequency and high frequency ratio of the heart rate is set to $0.5$, the angle, z-position, and Gaussian width are $-70$, $0$ and $0$ for the P wave ($-12\pm 2$, $-5\pm 10$, and $0.05\pm 0.01$  for the Q wave, $0$, $20\pm 8$, and $0.08\pm 0.01$ for the R wave, $12\pm 2$, $-15\pm 8$, and $0.07\pm 0.01$ for the S wave, and $90\pm 10$, $0.5\pm 0.2$, and $0.12\pm 0.02$ for the T wave respectively).

\section{Results}\label{Section:MaterialResult}

We apply the algorithm to a variety of real and simulated ECG signals, compare our algorithm to the previous algorithms described in Section \ref{Section:Other}, and use the evaluation metrics in Sections \ref{Section:metrics} and \ref{Section:metrics2} to measure the performance of different algorithms. We visually and quantitatively justify our use of the diffusion distance.  All of our experiments are conducted in Mathworks MATLAB R2015a using a 4th Generation Intel Core i7 processor and 24 GB of RAM.

\subsection{Comparison}

We validate $\ours$ and compare its performance with that of ABS \cite{Slocum1985,Slocum1992,Holm1998,Shah2004}, local PCA \cite{Castells2005}, non-local PCA, local aSVC \cite{Alcaraz2008} and non-local aSVC \cite{Alcaraz2008}. We also show the result with $\oursX$. When we run the local aSVC and non-local aSVC in the simulated signal, we use the ground truth heights of the Q and R points for the cancelation.

\begin{figure}
\centering
\includegraphics[width=.98\textwidth]{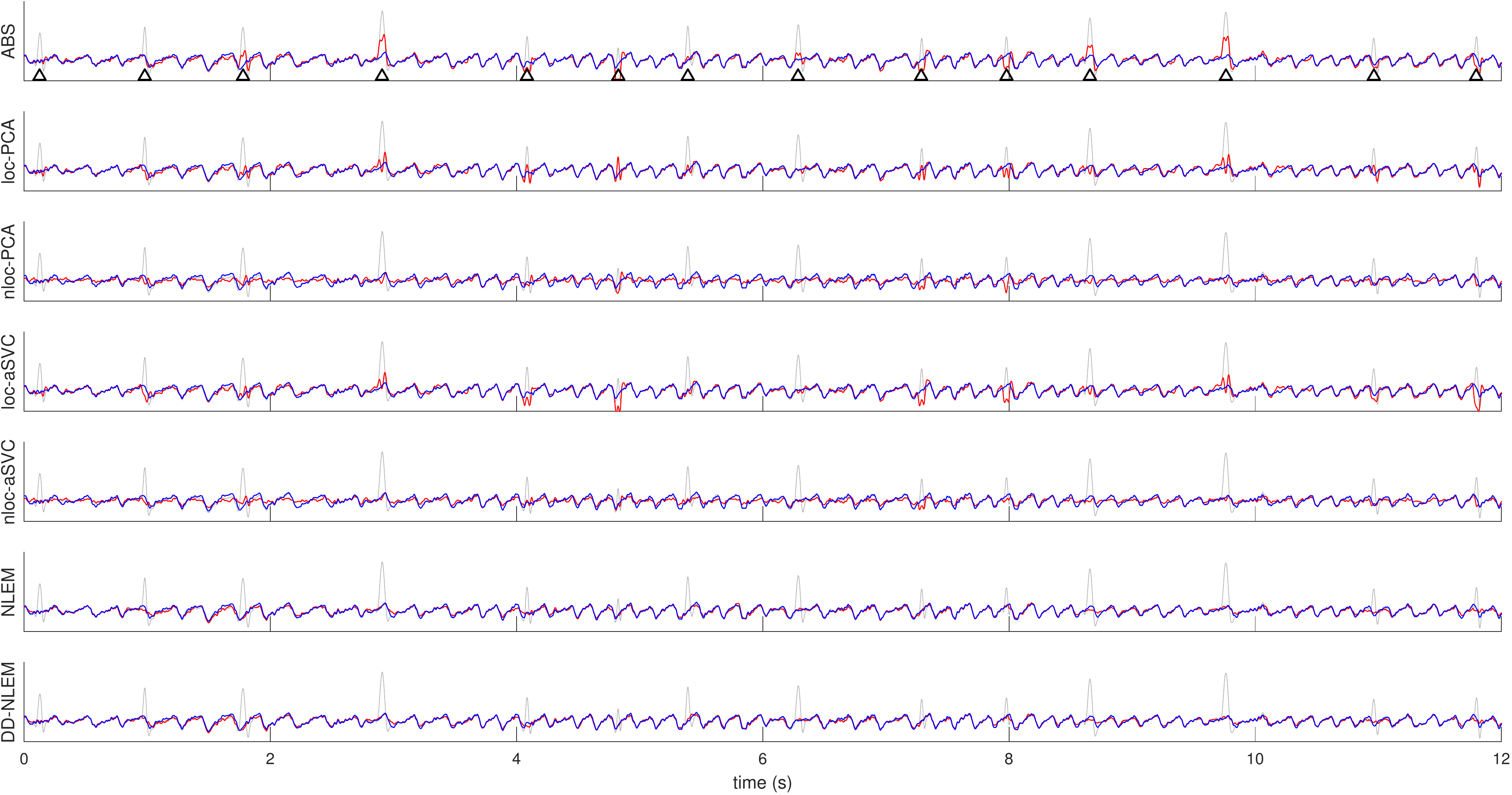}
\caption{The results of different f-wave extraction algorithms on the simulated f-wave by the sawtooth model. From top to bottom: average beat subtraction (ABS), local principal component analysis (PCA), non-local PCA, local adaptive singular value cancellation (aSVC), non-local aSVC, $\oursX$, and $\ours$. In the top row, the detected R peaks are marked by the black triangles. In each row, the simulated ECG recording is shown in the gray curve, the simulated f-wave is superimposed as a blue curve, and the extracted f-wave is superimposed as a red curve. It is clear that the outcomes of ABS and local aSVC have more ventricular residues, while in $\ours$ the residue is smaller. On the other hand, the non-local aSVC, as is discussed in the main text, leads to the over-smoothing effect since the f-wave is not pre-processed. \label{Numerics:SawResult}}
\end{figure}

We facilitate this comparison by generating three events.
In the first event, we evaluate the performance of our algorithm using synthesized signals generated by the sawtooth model, and the signal length is one hour.
We generate a total of 50 simulated signals and apply the above-mentioned algorithms to each signal.
Summary statistics of the results for the first event is shown in Table~\ref{Table:results2}. In this event, we see that both $\oursX$ and $\ours$ outperform other methods when evaluated by the considered evaluation indices, except the computational time, and $\ours$ performs the best. For the computational time, due to the nearest neighbor search, the non-local algorithms, like non-local PCA and non-local aSVC, take longer time than their local versions, like local PCA and local aSVC. On the other hand, $\oursX$ and $\ours$ are slightly more efficiently than the non-local PCA and non-local aSVC since we evaluate the median instead of the singular decomposition. 
The results of one typical sawtooth simulation by different algorithms are shown in Figure \ref{Numerics:SawResult}. 

\begin{figure}
\centering
\includegraphics[width=.98\textwidth]{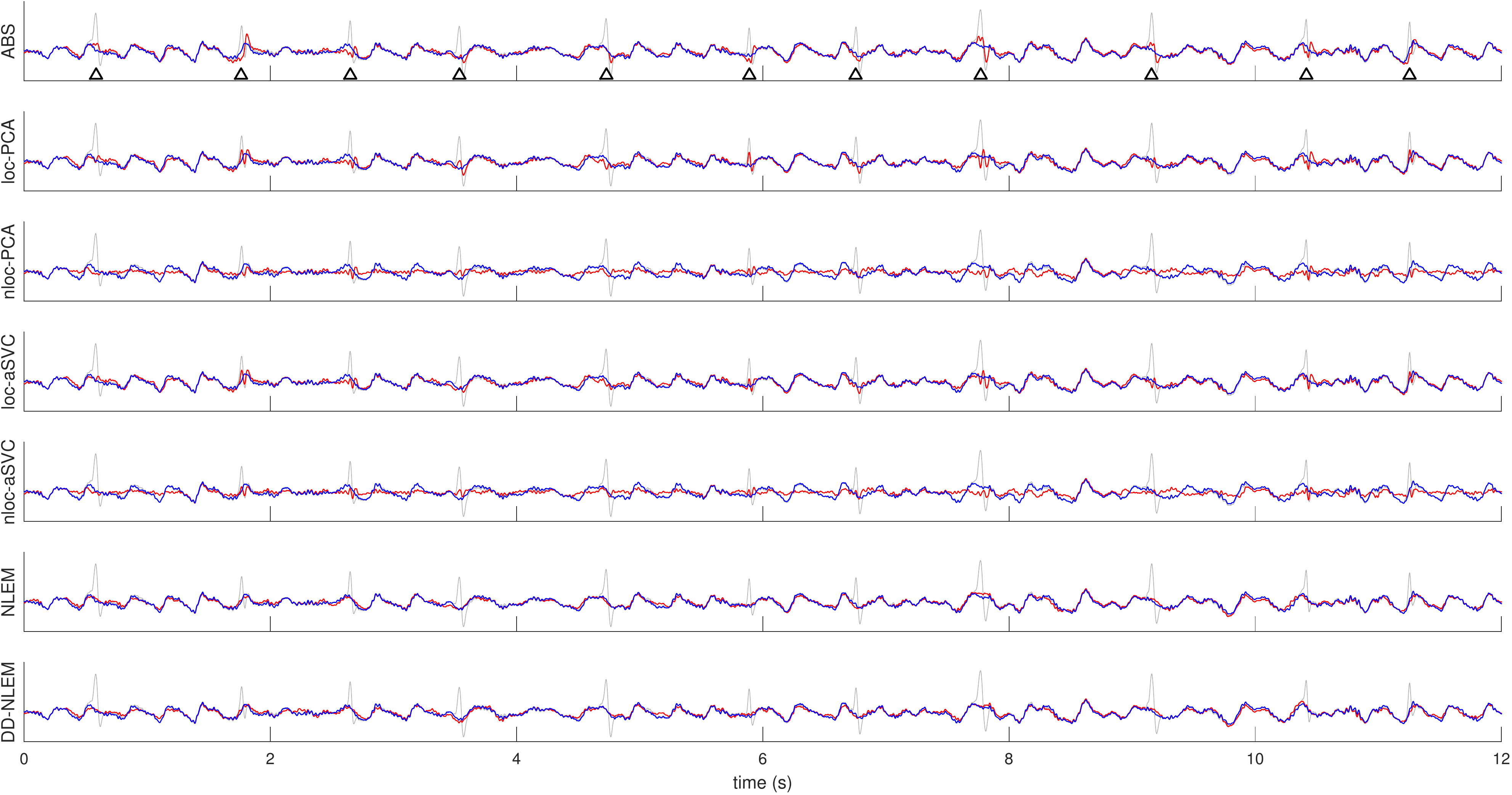}
\caption{The results of different f-wave extraction algorithms on the simulated f-wave by the excision method. From top to bottom: average beat subtraction (ABS), local principal component analysis (PCA), non-local PCA, local adaptive singular value cancellation (aSVC), non-local aSVC, $\oursX$, and $\ours$. In the top row, the detected R peaks are marked by the black triangles. In each row, the simulated ECG recording is shown in the gray curve, the simulated f-wave is superimposed as a blue curve, and the extracted f-wave is superimposed as a red curve. It is clear that the outcomes of ABS and local aSVC have more ventricular residues, while in $\ours$ the residue is smaller. On the other hand, the non-local aSVC, as is discussed in the main text, leads to the over-smoothing effect since the f-wave is not pre-processed. \label{Numerics:ExResult}}
\end{figure}

In the second event, we evaluate the performance of all algorithms using the signals that we generated by the excision method.
We synthesize 5 ventricular signals for each of the 9 excised f-wave signals and combine them for a total of 45 synthesized ECG signals. Each synthesized signal is one hour long.
The results of applying different algorithms to each of the 45 signals can be found in Table~\ref{Table:results22}. In this event, we see that both $\oursX$ and $\ours$ outperform other methods when evaluated by the considered evaluation indices, except the computational time, and $\ours$ performs better than $\oursX$ in the mVR metric.
In Figure \ref{Numerics:ExResult}, one typical simulation by the excision method of different algorithms is shown for the comparison.

\begin{figure}
\centering
\includegraphics[width=.98\textwidth]{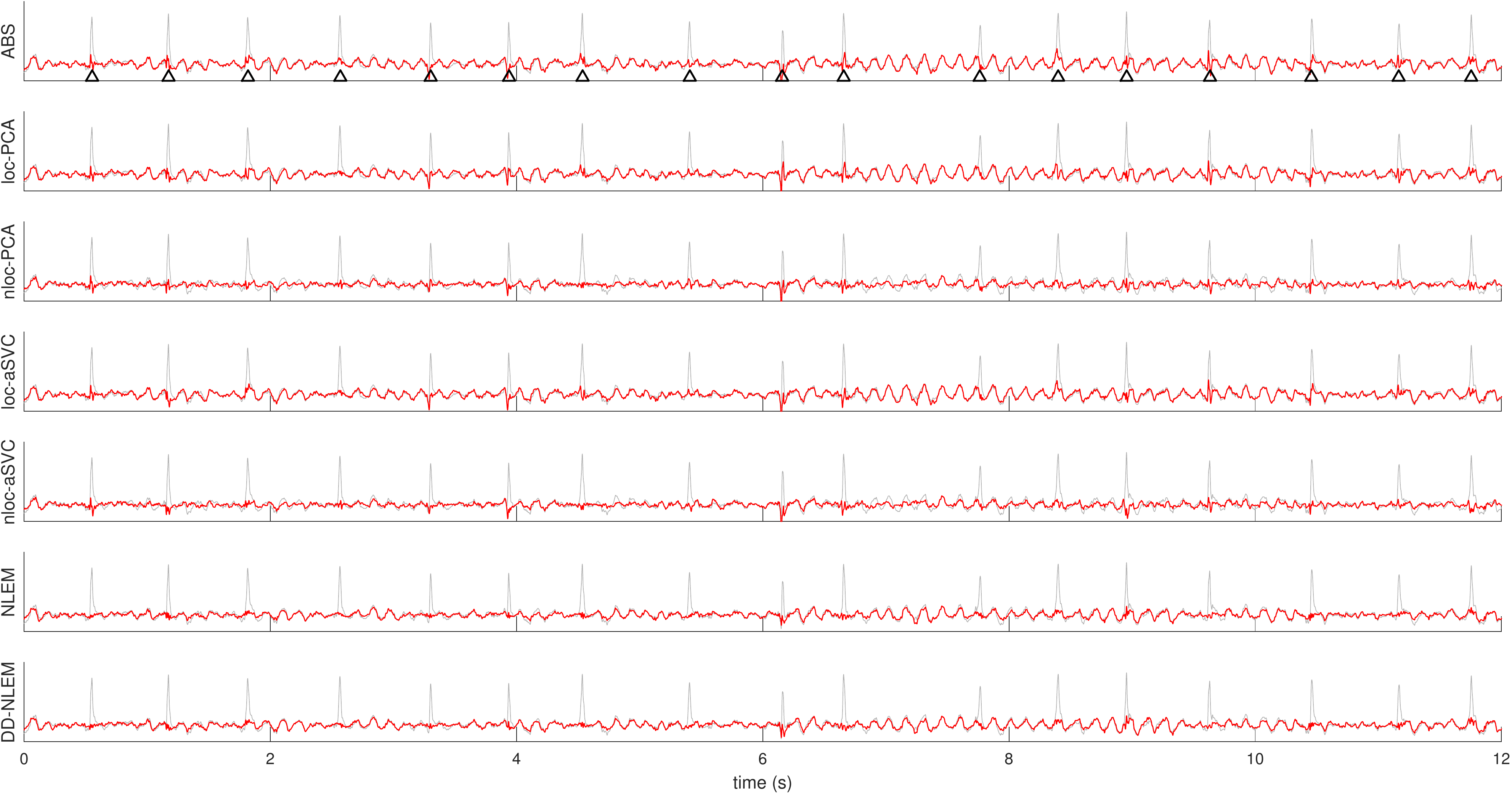}
\caption{The results of different f-wave extraction algorithms on a real holter recording. From top to bottom: average beat subtraction (ABS), local principal component analysis (PCA), non-local PCA, local adaptive singular value cancellation (aSVC), non-local aSVC, $\oursX$, and $\ours$. In the top row, the detected R peaks are marked by the black triangles. In each row, the original Holter recording is shown in the gray curve, and the extracted f-wave is superimposed as a red curve. It is clear that the outcomes of ABS and local aSVC have more ventricular residues, while in $\ours$ the residue is smaller. On the other hand, the non-local aSVC, as is discussed in the main text, leads to the over-smoothing effect since the f-wave is not pre-processed. \label{Numerics:HolterResult}}
\end{figure}

In the last event, we apply each of the 6 algorithms to 47 real Holter signals and display the corresponding VR and S indices values in Table~\ref{Table:resultsreal}. The proposed $\ours$ perform the best among all compared algorithms. The results of one typical subject of different algorithms are shown in Figure \ref{Numerics:HolterResult}.

Overall, we could see that the ABS, local PCA and local aSVC tend to have a better f-wave recovered outside the VA regions but with a larger ventricular remainder inside the VA regions compared with the non-local PCA and non-local aSVC. On the other hand, while the non-local PCA and and non-local aSVC remove the VA better, they tend to provide an ``over-smoothed'' f-wave estimation. One particular reason for the over-smoothness of the non-local PCA and non-local aSVC is that the f-wave is not handled before evaluating the cosine or correlation affinity between cardiac activities. When the signal is short, it could perform well. However, when the signal is long, which is the situation we encounter in the Holter signal or other monitoring devices, the neighboring cardiac activities determined by the cosine affinity will not only have similar VA, but also similar f-wave. As a result, the singular vector is contaminated by the f-wave, and could not serve as a good VA template. 
Another problem encountered in the PCA or SVC approaches is the {\em large $p$ and large $n$} issue, which has been widely discussed in the past decade \cite{Johnstone:2006}. Under the current setup, the f-wave behaves like a high dimensional ``random noise,'' while the VA could be viewed as the signal we want to recover. It has been known that under this large $p$ and large $n$ setup, the estimator of the VA would be biased, and a correction is needed, like \cite{Singer_Wu:2013a,Donoho_Gavish_Johnstone:2013}.
The problems encountered by these algorithms are taken care in the proposed $\oursX$ and $\ours$ -- the f-wave is pre-filtered before evaluating the $d_{\texttt{VS}}$, the DD is applied to reduce the possible error incurred in the whole procedure, and the median is evaluated to extract the VA.

\begin{table}
\scriptsize
\renewcommand{\arraystretch}{1.5}
\begin{tabular}{|c|c|c|c|c|c|c|c|}
\hline
Method & NMSE & $\rho$ & SNR & PSNR & time (s) & mVR & SC \\
\hline
ABS &  $0.44 \pm 0.22$  & $0.82 \pm 0.07$ & $4.19 \pm 2.51$   &  $16.34 \pm 2.39$ & $\mathbf{0.33 \pm 0.05}$ & $1.18 \pm 0.10$ & $0.68 \pm 0.15^{\dagger\ddagger}$ \\
\hline
local PCA & $0.33 \pm 0.11$ &$0.84 \pm 0.04$ &$4.99 \pm 1.38$  & $17.14 \pm 1.37$ & $7.87 \pm 1.07$ & $1.14 \pm 0.07$ & $0.63 \pm 0.16$ \\
\hline
non-local PCA & $0.56 \pm 0.17$ &$0.83 \pm 0.06$ &$2.69 \pm 1.37$  & $14.84 \pm 1.76$ & $34.53 \pm 7.65$ & $1.41 \pm 0.22$ & $0.61 \pm 0.06$ \\
\hline
local aSVC &  $0.39 \pm 0.14$ & $0.83 \pm 0.05$  & $4.38 \pm 1.62$    & $16.56 \pm 1.66$ & $17.29 \pm 3.30$ & $1.15 \pm 0.08$ & $0.65 \pm 0.18$ \\
\hline 
non-local aSVC &  $0.50 \pm 0.14$ & $0.72 \pm 0.11$  &  $3.21 \pm 1.22$ &   $15.36 \pm 1.63$ & $42.98 \pm 8.93$ & $1.38 \pm 0.20$ & $0.62 \pm 0.08$ \\
\hline
$\oursX$ & $0.18 \pm 0.05$ &  $0.91 \pm 0.03$ &  $7.66 \pm 1.23$  &  $19.81 \pm 1.34$ &  $16.13 \pm 3.61$ & $1.09 \pm 0.03$ & $0.68 \pm 0.18^\ddagger$ \\
\hline
$\ours$ &  $\mathbf{0.17 \pm 0.03}$ & $\mathbf{0.91 \pm 0.02}$ & $\mathbf{7.71\pm 0.79}$ & $\mathbf{19.86 \pm 0.87}$ & $21.73 \pm 4.61$ & $\mathbf{1.08 \pm 0.01}$ & $\mathbf{0.69 \pm 0.18}^\dagger$ \\
\hline
\end{tabular}
\caption{Simulation results for extraction of sawtooth f-waves. The results over all simulations are summarized as mean $\pm$ the standard deviation. For each evaluation metric, the best method is marked in the boldface. 
{For each previous method (ABS, local PCA, non-local PCA, local aSVC, and non-local aSVC), we test the null hypothesis that the errors  between the previous method and $\ours$ (or $\oursX$) is different with the one-sided sign test.  
A p-value less than 0.05 was considered as statistically significant, and multiple significant tests were calculated with the Bonferroni correction. $\dagger$: the p-value is {\em not} less than 0.05 between each previous method and $\ours$; $\ddagger$: the p-value is {\em not} less than 0.05 between each previous method and $\oursX$.}
}\label{Table:results2}
\end{table}


\begin{table}
\scriptsize
\renewcommand{\arraystretch}{1.5}
\begin{tabular}{|c|c|c|c|c|c|c|c|}
\hline
Method & NMSE & $\rho$ & SNR & PSNR & time (s) & mVR & SC \\
\hline
ABS &  $1.55 \pm 1.83$  & $0.67 \pm 0.15$ &               $-0.10 \pm 3.70$   &  $13.65 \pm 3.57$ & $\mathbf{0.33 \pm 0.04}$ & $1.49 \pm 0.54$ & $0.66 \pm 0.05$ \\
\hline
local PCA & $0.88 \pm 0.89$ & $0.73 \pm 0.13$ & 		$1.83 \pm 3.12$ & $15.57 \pm 3.05$ & $7.85 \pm 0.88$& $1.34 \pm 0.37$& $0.57 \pm 0.10$\\
\hline
non-local PCA & $0.35 \pm 0.05$ & $0.81 \pm 0.03$ & 	$4.60 \pm 0.61$ & $18.35 \pm 1.02$ & $33.89 \pm 6.85$& $1.10 \pm 0.03$& $0.58 \pm 0.02$\\
\hline
local aSVC &  $1.33 \pm 1.41$ & $0.67 \pm 0.15$  &        $0.34 \pm 3.6$  & $14.08 \pm 3.54$ & $16.36 \pm 1.94$ & $1.44 \pm 0.53$ & $0.60 \pm 0.08$ \\
\hline 
non-local aSVC &  $0.34 \pm 0.05 $ & $0.82 \pm 0.03$  &  $4.74 \pm 0.55$ &   $18.55 \pm 0.82$ & $42.34 \pm 8.13$ & $1.10 \pm 0.03$ & $0.58 \pm 0.02$ \\
\hline
$\oursX$ & $\mathbf{0.24 \pm 0.04}$ &  $\mathbf{0.87 \pm 0.02}$ & $\mathbf{6.28 \pm 0.63}$  &  $\mathbf{20.03 \pm 0.88}$ &  $15.51 \pm 3.39$ & $1.07 \pm 0.01$ & $\mathbf{0.72 \pm 0.03}$ \\
\hline
$\ours$ &  $0.25 \pm 0.05$ & $0.87 \pm 0.03$ &          	$6.14 \pm 0.81$ & $19.89 \pm 1.01$ & $21.16 \pm 4.18$ & $\mathbf{1.06 \pm 0.00}$ & $\mathbf{0.72 \pm 0.03}$ \\
\hline
\end{tabular}
\caption{Simulation results for extraction of excised f-waves. The results over all simulations are summarized as mean $\pm$ the standard deviation. For each evaluation metric, the best method is marked in the boldface.
{For each previous method (ABS, local PCA, non-local PCA, local aSVC, and non-local aSVC), we test the null hypothesis that the errors  between the previous method and $\ours$ (or $\oursX$) is different with the one-sided sign test.  
A p-value less than 0.05 was considered as statistically significant, and multiple significant tests were calculated with the Bonferroni correction. $\dagger$: the p-value is {\em not} less than 0.05 between each previous method and $\ours$; $\ddagger$: the p-value is {\em not} less than 0.05 between each previous method and $\oursX$.}
}\label{Table:results22}
\end{table}


\begin{table}
\centering
\scriptsize
\renewcommand{\arraystretch}{1.5}
\begin{tabular}{|c|c|c|c|}
\hline
Method & mVR & SC & time (s)\\
\hline
ABS &   $1.56 \pm 0.49$  &  $0.51 \pm 0.12^{\dagger\ddagger}$   & $\mathbf{0.42 \pm 0.10}$\\
\hline
local PCA & $1.56 \pm 0.47$ & $0.50 \pm 0.11^\ddagger$ & $9.93 \pm 2.23$ \\
\hline
non-local PCA & $1.35 \pm 0.45$ & $0.41 \pm 0.09$ & $52.79 \pm 23.56$\\
\hline
local aSVC & $1.92 \pm 0.81$    &  $0.50 \pm 0.12$ & $22.50 \pm 5.72$\\
\hline
non-local aSVC &  $1.39 \pm 0.32$   & $0.46 \pm 0.06$  & $64.32 \pm 26.81$ \\
\hline
$\oursX$ & $1.25 \pm 0.59$   & $0.51 \pm 0.14^\ddagger$ & $24.92 \pm 11.30$\\
\hline
$\ours$ &  $\mathbf{1.21 \pm 0.35}$   &  $\mathbf{0.52 \pm 0.14}^\dagger$ & $32.85 \pm 13.95$\\
\hline
\end{tabular}
\caption{Evaluation metrics for real Holter signals.  The results over all subjects are summarized as mean $\pm$ the standard deviation.  For each evaluation metric, the best method is marked in the boldface.
{For each previous method (ABS, local PCA, non-local PCA, local aSVC, and non-local aSVC), we test the null hypothesis that the errors  between the previous method and $\ours$ (or $\oursX$) is different with the one-sided sign test.  
A p-value less than 0.05 was considered as statistically significant, and multiple significant tests were calculated with the Bonferroni correction. $\dagger$: the p-value is {\em not} less than 0.05 between each previous method and $\ours$; $\ddagger$: the p-value is {\em not} less than 0.05 between each previous method and $\oursX$.}
}\label{Table:resultsreal}
\end{table}


\subsection{The validation of the proposed mVR index}

\begin{figure}
\centering
\includegraphics[width=.45\textwidth]{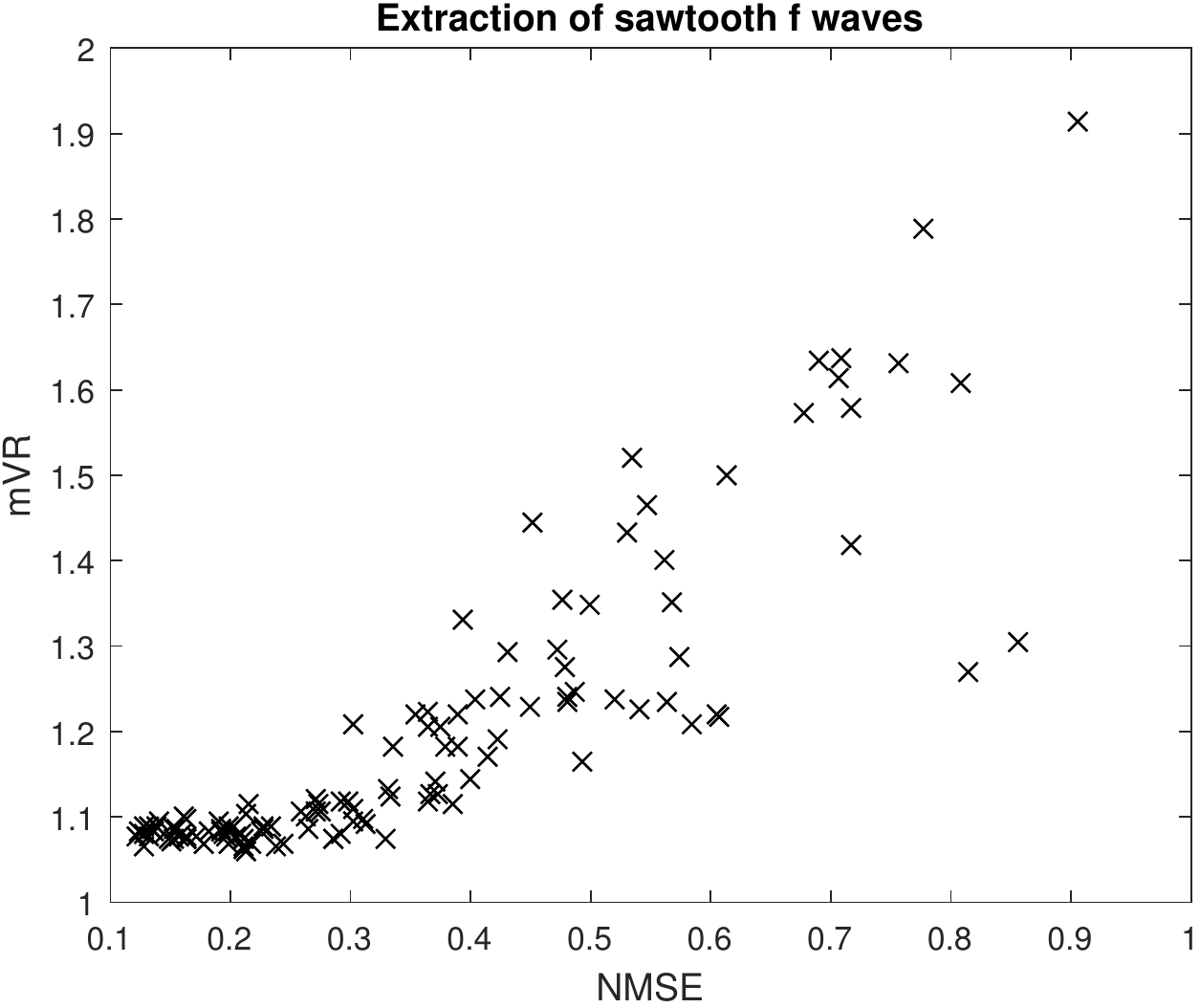}
\includegraphics[width=.45\textwidth]{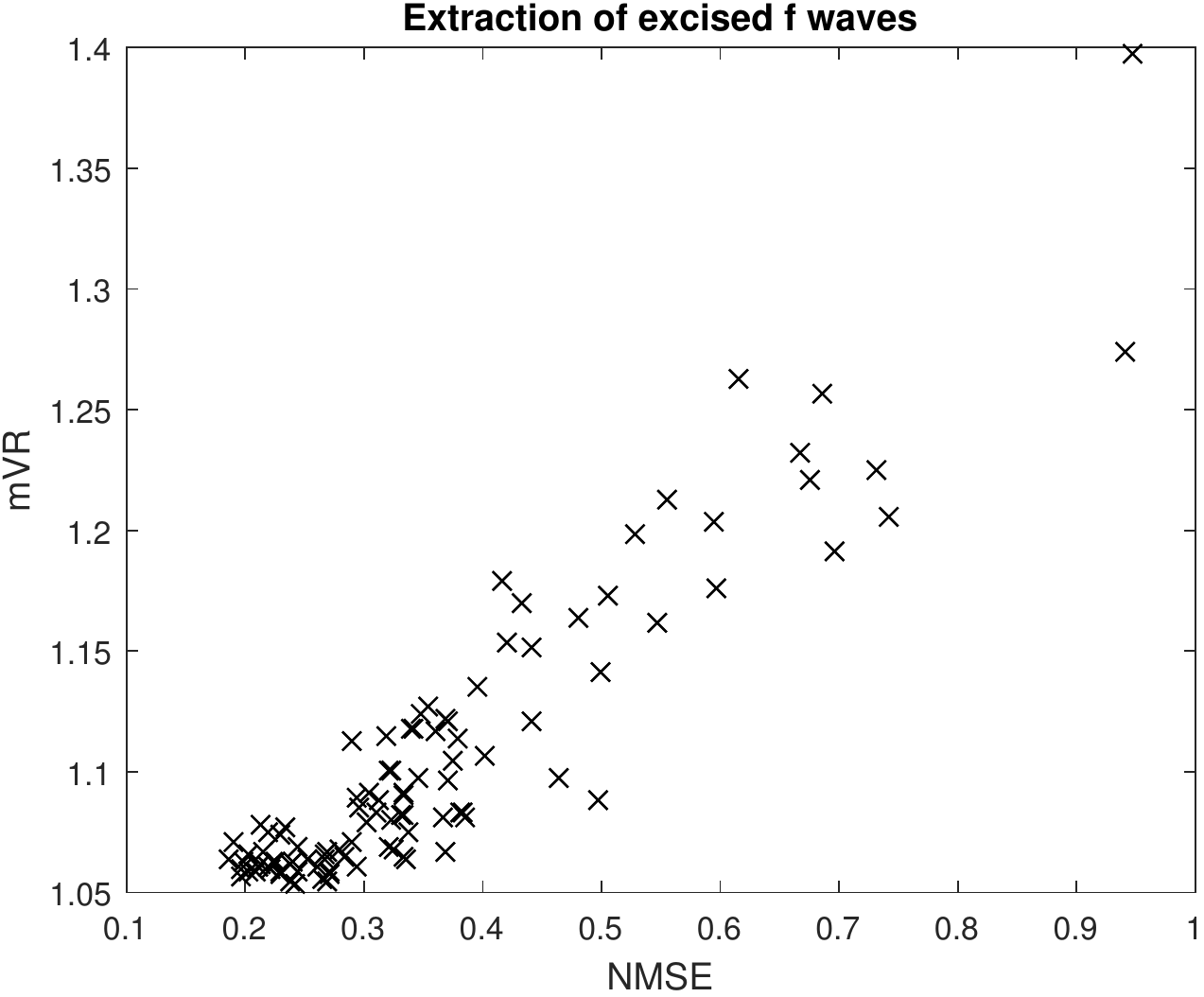}
\caption{The scatter plot of the mVR index versus the normalized mean square error (NMSE). Left: the results of all simulated signals generated by the sawtooth model by seven different f-wave extraction algorithms; right: the results of all simulated signals generated by the excision method by seven different f-wave extraction algorithms. \label{Numerics:mVRvsNMSE}}
\end{figure}

To validate the usefulness of the proposed mVR index in the real signal, we examine carefully the relationship between the mVR index and the NMSE index in the simulated signal. In Figure \ref{Numerics:mVRvsNMSE}, the scatter plot of the mVR versus the NMSE of 7 algorithms over all simulated signals are shown. It is clear that the order of mVR and the order of NMSE are correctly related when NMSE is greater than 0.3. This indicates that while mVR might be insensitive to the small error, mVR could well catch big errors. In addition to the philosophy behind the design of mVR discussed above, these results support the usefulness of the proposed mVR index when applied to the real data.

We demonstrate one example to compare the VR and mVR indices. We generate one simulated ECG signal and run non-local aSVC and $\ours$ to extract the f wave. In Figure \ref{fig:VRindexcomparison}, we take a few beats and assess their VR and mVR indices. Visually we see that $\ours$ well extracts the f-wave, while we observe significant over-smoothing of the f wave after non-local aSVC. However, the VR index is insensitive to this over-smoothing and suggests a better result by non-local aSVC. By taking a closer look, we see that the VR index is sensitive to the amplitude of the f wave underneath each QRS complex, and this could generate the instability of the VR index.  On the other hand, the mVR index is capable of identifying both unwanted ventricular residua and the quality of f wave preservation.

\begin{figure}

\centering
\includegraphics[width=.9\textwidth]{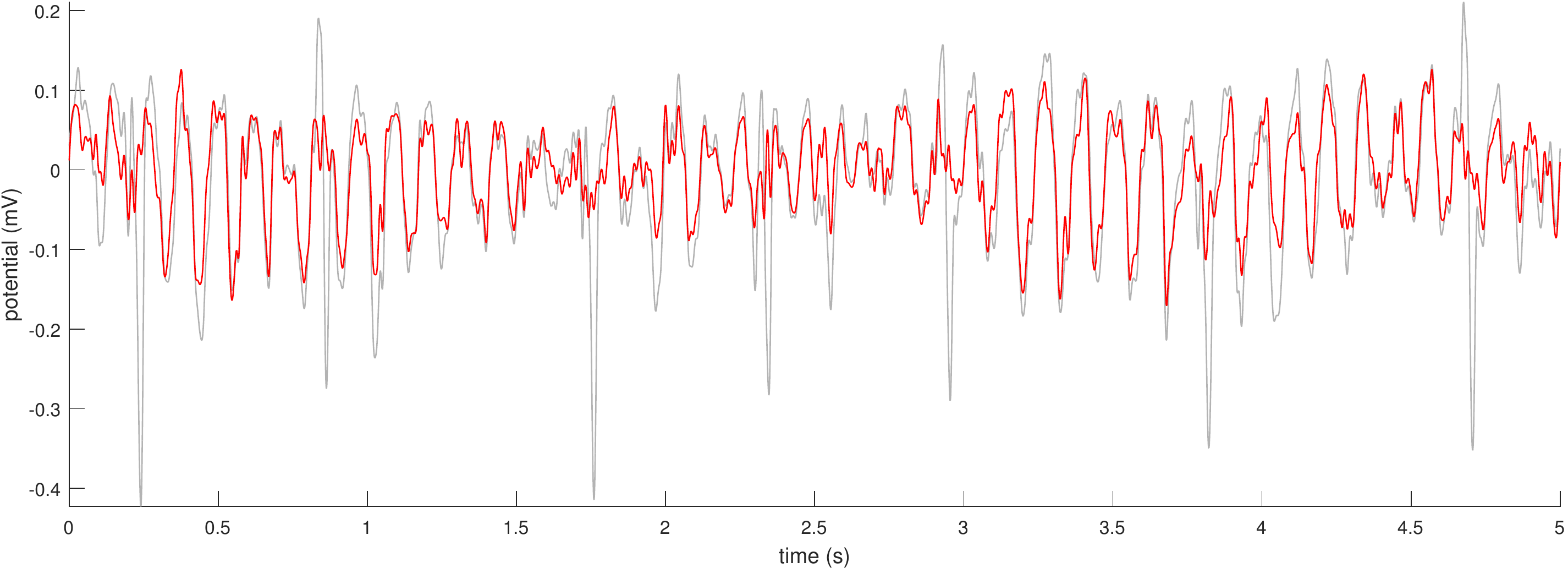}
\caption{{A real Holter signal from a patient with persistent atrial fibrillation recorded from channel 1, E(+) to S(-), of DigiTrak XT (Philips) under the EASI lead system. The original Holter recording is shown in the gray curve, and the extracted f-wave is superimposed as a red curve. Clearly, the f wave is significantly large compared with the ventricular response.} \label{fig:RealBigFwaveResult}}
\end{figure}

{At the first glance, the size of the simulated f-waves in Figure \ref{fig:VRindexcomparison} seems unrealistically large compared to the QRS. It is however not unrealistic; from time to time it could be observed in the real Holter signal. See Figure \ref{fig:RealBigFwaveResult} for a real Holter signal example.
Previous non-local algorithms struggle the most when the amplitude of the f wave is high, and Figure \ref{fig:VRindexcomparison} is meant in part to emphasize this weakness. It is also worth mentioning that we have not received the impression that previous works account for these drastic cases, and we would like to be clear that we do account for them.
Lastly, the f wave amplitude in Figure \ref{fig:VRindexcomparison} is chosen to be large because the VR index in \ref{Definition:VRindex} will be inappropriately large when the f wave extraction is performed accurately; that is, the old VR index rewards the extracted f wave if it has low amplitude underneath the QRS complex.}

\tikzstyle{line} = [draw, -latex']
\tikzstyle{arrow} = [thick,->,>=stealth]
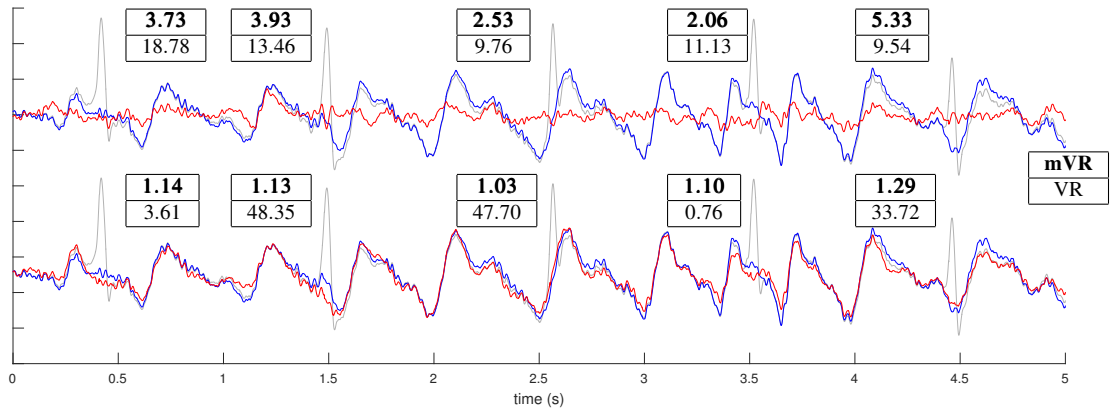
\begin{figure}
\begin{tikzpicture}[>=latex']
        \tikzset{block/.style= {draw, rectangle, align=center,minimum width=2cm,minimum height=.7cm,line width=0.3mm},
        lblock/.style={draw, rectangle, align=left,minimum width=2cm,minimum height=.7cm,line width=0.3mm},
        smallblock/.style={draw, rectangle, rounded corners = 1, align=center,minimum width=2cm,minimum height=.7cm,line width=0.18mm},
        input/.style={ 
        draw,
        trapezium,
        trapezium left angle=60,
        trapezium right angle=120,
        minimum width=2cm,
        align=center,
        minimum height=.7cm
    },
        }

        \node [text width=0.9\textwidth] (graphic) {\includegraphics[width=1.0\textwidth]{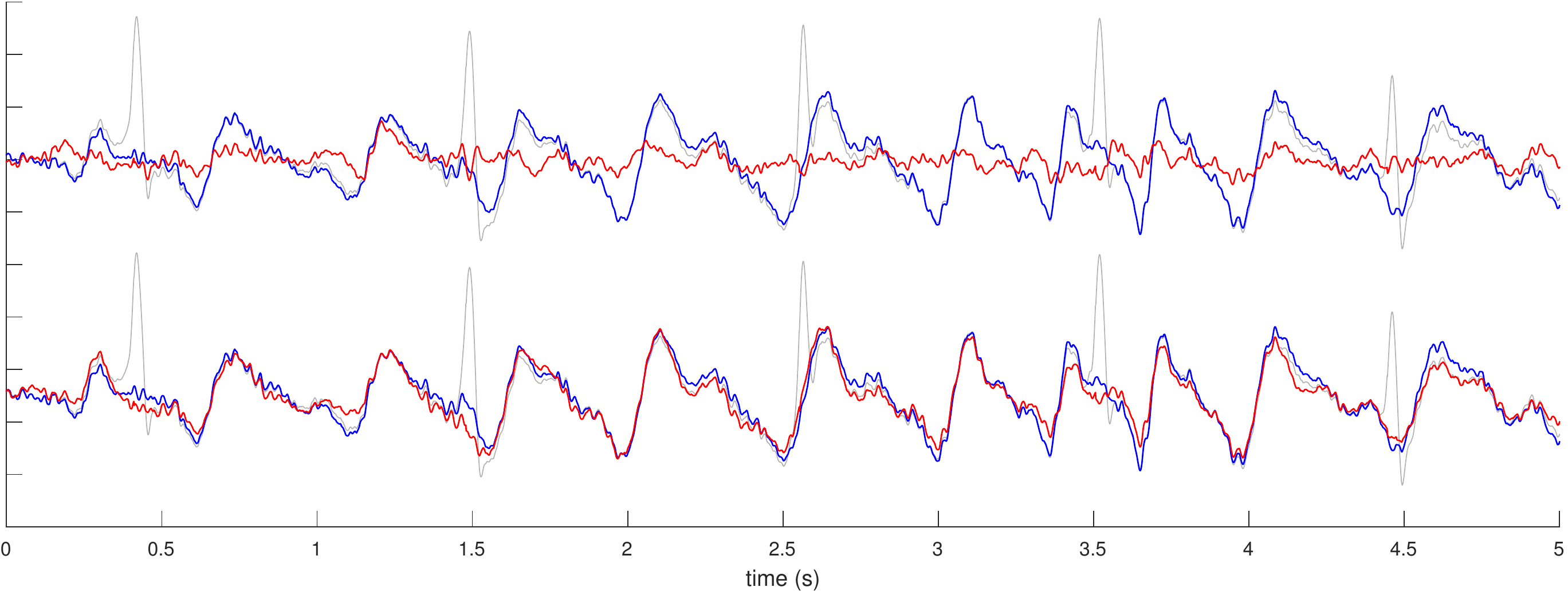}};
        \node [text width=3cm] at (-4.0, 2.3) (graphich) {\scriptsize  \begin{tabular}{|c|} \hline $\boldsymbol{3.73}$ \\ \hline $18.78$ \\ \hline \end{tabular}};
        \node [text width=3cm] at (-2.6, 2.3) (graphici) {\scriptsize  \begin{tabular}{|c|} \hline $\boldsymbol{3.93}$ \\ \hline $13.46$ \\ \hline \end{tabular}};
        \node [text width=3cm] at (0.4, 2.3) (graphicj) {\scriptsize  \begin{tabular}{|c|} \hline $\boldsymbol{2.53}$ \\ \hline $9.76$ \\ \hline \end{tabular}};
        \node [text width=3cm] at (3.2, 2.3) (graphick) {\scriptsize  \begin{tabular}{|c|} \hline $\boldsymbol{2.06}$ \\ \hline $11.13$ \\ \hline \end{tabular}};
        \node [text width=3cm] at (5.7, 2.3) (graphicl) {\scriptsize  \begin{tabular}{|c|} \hline $\boldsymbol{5.33}$ \\ \hline $9.54$ \\ \hline \end{tabular}};
        
        \node [text width=3cm] at (-4.0, 0.1) (graphich) {\scriptsize  \begin{tabular}{|c|} \hline $\boldsymbol{1.14}$ \\ \hline $3.61$ \\ \hline \end{tabular}};
        \node [text width=3cm] at (-2.6, 0.1) (graphici) {\scriptsize  \begin{tabular}{|c|} \hline $\boldsymbol{1.13}$ \\ \hline $48.35$ \\ \hline \end{tabular}};
        \node [text width=3cm] at (0.4, 0.1) (graphicj) {\scriptsize  \begin{tabular}{|c|} \hline $\boldsymbol{1.03}$ \\ \hline $47.70$ \\ \hline \end{tabular}};
        \node [text width=3cm] at (3.2, 0.1) (graphick) {\scriptsize  \begin{tabular}{|c|} \hline $\boldsymbol{1.10}$ \\ \hline $0.76$ \\ \hline \end{tabular}};
        \node [text width=3cm] at (5.7, 0.1) (graphicl) {\scriptsize  \begin{tabular}{|c|} \hline $\boldsymbol{1.29}$ \\ \hline $33.72$ \\ \hline \end{tabular}};
        
        \node [text width=3cm] at (8.0, 0.4) (graphicn) {\scriptsize  \begin{tabular}{|c|} \hline \textbf{mVR} \\ \hline VR \\ \hline \end{tabular}};
       
\end{tikzpicture}
\caption{Assessment of VR and mVR indices for two extracted f waves for the simulated signal. The simulated ECG signal is plotted in gray, the simulated f-wave is superimposed as the blue curve, and the extracted f-waves are superimposed as the red curves, where the non-local aSVC result is shown in the top and the result by $\ours$ is shown in the bottom. It is visually clear that the extracted f-wave by $\ours$ is closer to the simulated f-wave, however, the VR index provides an opposite information.} \label{fig:VRindexcomparison}
\end{figure}

\section{Discussion}\label{Section:Discussion}

Designing a ``correct'' metric is the core of data analysis. In this work, our metric design process involves two steps. First, inspired by the Elgendi's QRS detection algorithm \cite{Elgendi2013}, based on the physiological knowledge of the f-wave, we take the morphological and spectral characteristics of VA and design the ventricular similarity metric over the patch space $\mathcal{X}$. However, the ventricular similarity metric is error-prone due to the intricate interplay between the VA and the f-wave. We thus apply the recently developed DM and DD to stabilize the ventricular similarity metric. We call the final metric the diffusion ventricular similarity metric. The DM and DD has been theoretically studied and shown to be robust to the possibly existing error in the affinity estimation.

The proposed $\ours$ has two novelties over traditional single-lead f-wave extraction algorithms. First, the non-local nature of the algorithm allows us to leverage information from the entire signal to robustly and accurately estimate a template for the ventricular activity in each beat; second, the designed ventricular similarity metric prevents the possible overfitting issue. For the first novelty, in addition to the theoretical justification of the diffusion ventricular similarity metric, compared with the ABS, local PCA and local aSVC approach, the experimental results support that the morphology of the VA is better preserved in $\ours$. For the second novelty, note that the non-local PCA and non-local aSVC algorithms have the over-smoothing issue. Specifically, since the cosine or correlation affinity is applied directly to the patch space, when the signal is long, the determined similar beats will have not only similar VA, but also similar f-wave. Thus, applying these algorithms directly to a long signal would lead to an underestimated f-wave. Based on these two novelties, $\ours$ is suitable for analyzing long term signal with high accuracy. 

There are several topics regarding the single-lead f-wave analysis not touched in this paper. The most important one is the outlier issue. Specifically, when there exist beats that are far away from all other beats in terms of the ventricular similarity metric or the diffusion ventricular similarity metric, the algorithm may not perform well. The commonly encountered ectopic beat is an example. The existence of ectopic beats poses a unique problem in a number of ways. They vary significantly in morphology when compared to the signal's normal beats. They can also vary significantly among themselves, and the number of ectopic beats in a 1-hour recording may be as low as $1$-$2$. As such, local ABS-type algorithm is severely limited, and the non-local nature algorithm is needed. In this work, we propose to apply the proposed algorithm in \cite{Martinez2013} to separate the ectopic beats, and cancel the VA separately. While it gives a reasonable performance in the real database, it still has a room to improve the algorithm. For example, we could consider further clustering the ectopic beats before running the non-local cancellation algorithm, or we could pool the ectopic beats from different subjects to catch more accurately the morphological features of an ectopic beat. 
{If the number of ectopic beats is exceptionally small, like smaller than $k_0$, then the result would be largely unaffected by these errors. In this situation, we recommend leaving these beats ``as is,'' or running the cancellation algorithm with the only available ectopic beats.}
Another issue is the inevitable noise. Note that even if $\ours$ or other single-lead f-wave extraction algorithms could accurately get the VA template, the inevitable noise still exists and we may need post-processing, depending on the application. There are several possibilities for this purpose, for example, adaptive recurrent filtering algorithms adapted to successive beats for subtraction \cite{Sahakian1985}. In this paper, to simplify the discussion, we do not incorporate this step.
The T wave is not handled separately in this work. It is handled jointly with the QRS complex under the assumption that the T wave and the QRS complex are related. However, there have been some works in the field showing the benefit of dealing with the T wave separately, like \cite{Lemay2005}, and this suggests the potential to incorporate the T wave cancellation in $\ours$. 
These topics will be explored and reported in the future.

We mention several future works. 
First, in this work, we focus on the persistent Af patients. The analysis for the paroxysmal Af patients is of its importance and will be explored in the upcoming research, for example, how to detect the spontaneous initiation and termination of the Af.
Second, while we apply the fiber bundle structure to study the cardiac activities, the topological structure of the fiber bundle is not fully utilized. We will focus on the theoretical exploration of the topological information. On the other hand, we could directly apply the recently proposed horizontal diffusion map \cite{Gao:2016} to further explore the fiber bundle. 
Third, once the f-wave signal is obtained, we need an informative feature representing the atrial electrophysiological properties from the extracted f-wave \cite{Lankveld2014,Bonizzi2014}, so that these features could reflect the patient's disease status for the diagnosis, treatment, and prognosis purposes.
There have been several indices proposed to quantify the intrinsic features underlying the f-wave, for example, the effect of electrophysiological remodeling, the nature of the fibrillation \cite{Hoppe2005}, to name but a few. 
See \cite{Lankveld2014,Bonizzi2014} for a summary of the available indices. While these indices are informative for different clinical applications \cite{Lankveld2016}, the time-varying dynamics hidden inside the f-wave, which is local in nature, are not fully extracted. We will apply other non-linear algorithms to extract these dynamical features and study clinical problems.
Fourth, although we confirm the performance of our algorithm based on simulated data, it is still not real. Since it is now relatively easy to obtain the intracardiac AA signal, we plan to study the relationship between the f-wave in the surface ECG and the intracardiac AA signal, and this relationship would allow us to better understand the f-wave and hence design a better algorithm to decouple the f-wave from the single-lead recording.

\section{Acknowledgement}
Hau-tieng Wu's research is partly supported by Sloan Research Fellow FR-2015-65363. Chun-Li Wang's research is partly supported by grants CMRPG3E2101 from the Chang Gung Memorial Hospital, Linkou, Taiwan.

\bigskip

\bibliographystyle{amsplain} 
\bibliography{fwave}

\clearpage 
\appendix

\renewcommand{\thefigure}{SI.\arabic{figure}}
\setcounter{figure}{0}
\renewcommand{\thetable}{SI.\arabic{table}}
\setcounter{table}{0}
\renewcommand{\theequation}{S.\arabic{equation}}
\setcounter{equation}{0}
\setcounter{page}{1}
\renewcommand{\thepage}{SI.\arabic{page}}

{\Large\center Online Supplementary Information for \\
\LARGE \center \textbf{Single-lead f-wave extraction using diffusion geometry}\\
\large\center by John Malik, Neil Reed, Chun-Li Wang, Hau-tieng Wu\\}

\bigskip\bigskip

\section{Mathematical background}\label{Section:Model}

\subsection{Mathematical model for the single-lead f-wave extraction algorithm}

We start the model by recalling the {\em vectocardiogram} (VCG) and its relationship with the surface ECG signals.
Denote the 12-lead surface ECG signals as $E(t)\in \RR^{12}$, which are the projection of the representative dipole current of the electrophysiological cardiac activity on different directions \cite{Keener:1998}. The discussion below holds for the other ECG setups, like body surface potential map \cite{Zeemering2014}, but to simplify the discussion, we focus on the commonly used 12-lead surface ECG signals. Denote the representative dipole current as $d(t)\in\RR^3$, where $t\in\RR$, which is in general called the VCG signal. Physiologically, $d(t)$ is oscillatory, and the period changes from time to time like a noise for a subject with atrial fibrillation (Af) \cite{ZengGlass1996}. The VCG signal is composed of the atrial activity (AA) and ventricular activity (VA), denoted as $a(t)\in\RR^3$ and $v(t)\in\RR^3$ respectively; that is, $d(t)=a(t)+v(t)$. For a normal subject, $a(t)$ and $v(t)$ are both periodic with the period of length about 1 sec, and ``synchronize'' perfectly. However, for a subject with Af without any pathological change in the ventricle, the dynamical behavior of the atrial and ventricular should be further discussed. First, the atrial activity $a(t)$ satisfies
\begin{enumerate}
\item it may or may not be periodic, depending on the type \cite{Hoppe2005};
\item the periodicity of $a(t)$ is much shorter (the fibrillation frequency could be {about 350 to 600 beats per minute \cite{Goodacre2002}}).   
\end{enumerate}
On the other hand, the ventricular activity $v(t)$ satisfies 
\begin{enumerate}
\item $v(t)$ is periodic with the period changing from time to time like a noise;
\item the oscillation of $v(t)$ dominates the oscillation of $d(t)$; 
\item the oscillation of $v(t)$ is not synchronized with $a(t)$; 
\item the ventricular responses are similar from time to time;
\item the length of the ventricular response is non-linearly related to the R peak to R peak interval.
\end{enumerate}
Here, based on the electrophysiology of cardiac activity, we know that unless under a pathological ventricular condition, the support of each ventricular response does not overlap. Precisely, we set the following {\em almost identical assumption}:

\begin{Assumption}\label{Condition:ECG}
Denote $I_k=[b_k,e_k]\subset \RR$ to be the time interval of the $k$th depolarization of ventricular response, where $b_k<t_k<e_k$ and $b_k$ (respectively $t_k$ and $e_k$) is the timing of the beginning of the ventricular depolarization (respectively, the R peak and the end of the ventricular depolarization). Then, there exist $s_k,a_k>0$ and $R_k\in O(3)$, $k\in\NN$, so that for all $k\neq l$
\begin{enumerate}
\item $I_k\cap I_l=\emptyset$,
\item $e_k-b_k$ is non-linearly related to $t_{k+1}-t_k$,
\item $\frac{1}{a_k}R_k^{-1}d\big(t_k+\frac{\delta}{s_k}\big)\sim \frac{1}{a_l}R_l^{-1}d\big(t_l+\frac{\delta}{s_l}\big)$, where $\delta\in\RR$ satisfies $t_k+\delta/s_k\in I_k$ and $t_l+\delta/s_l\in I_l$.    
\end{enumerate}
Here, $s_k$ describes the dilation of the cardiac activity, $a_k$ describes the amplitude of the cardiac activity, and $R_k$ describes the cardiac axis, and $\sim$ is commonly understood as the $L^2$ norm sense.
\end{Assumption}

For the $\ell$th surface ECG signal, where $\ell=1,\ldots,12$, there is an associated projection direction $u_\ell\in \RR^3$, and the $\ell$th ECG channel is the projection of $d(t)$ on $u_\ell$; that is, 
\begin{equation}
E_\ell(t)=u_\ell^Td(t)
\end{equation}
or $E(t)=u^Td(t)$, where $u=[u_1\,\,u_2 \ldots u_{12}]\in \RR^{3\times 12}$. Denote 
\begin{equation}
v_\ell(t)=u_\ell^Tv(t)\mbox{ and }a_\ell(t)=u_\ell^Ta(t).
\end{equation} 
Note that in general, $u_\ell$ changes according to time due to the cardiac axis deviation caused by the respiratory activity and other physical movements. To simplify the discussion, here we assume that $u_\ell$ is fixed, and we also assume that $R_k$ is the same for all $k\in\mathbb{N}$. Since the projection operator is continuous, the ventricular responses, normally shown as the QRST complex in the surface ECG signal, which is represented in $v_\ell$, are similar from time to time, while the similarity is further deformed by the cardiac axis deviation. In other words, Condition \ref{Condition:ECG}(3) further becomes
$E_\ell(t_k+\delta/s_k)/a_k\sim E_\ell(t_l+\delta/s_l)/a_l$, where $\ell=1,\ldots,12$ and $\delta\in\RR$ satisfies $t_k+\delta/s_k\in I_k$ and $t_l+\delta/s_l\in I_l$.

Based on Assumption \ref{Condition:ECG}, we have the following {\em phenomenological model}, which essentially captures the physiological fact that the VA and the AA are independent in the Af patient. Here, again to simply the discussion, we assume that $I_k$ are of the same length and contains the support of the VA. 

\begin{Assumption}
$\mathcal{V}_\ell:\{v_\ell|_{I_k}\}_{k\in\NN}\subset L^2(I)$ are sampled independently and identically (i.i.d.) from a function-valued random vector $\mathbf{V}$ and $\mathcal{A}_\ell:=\{a_\ell|_{I_k}\}_{k\in\NN}\subset L^2(I)$ are i.i.d. sampled from a function-valued random vector $\mathbf{A}$, where $|I|=|I_k|$ for all $k\in\mathbb{N}$. Here, $\mathbf{A}$ and $\mathbf{V}$ are independent.
\end{Assumption}

The goal of the single-lead f-wave extraction algorithm is to cluster the set of cardiac activities $\mathcal{E}_\ell=\{E_\ell|_{I_k}\}_{k\in\NN}=\{v_\ell|_{I_k}+a_\ell|_{I_k}\}_{k\in\NN}$ according to the similarity in $\mathcal{V}_\ell$, so that the VA inside a single clustered group is the same. Thus, based on the assumption that the f-wave is independent from the VA, we could count on a well-chosen ``averaging'' process to recover the VA.
However, in practice we only have $\mathcal{E}_\ell$ but not $\mathcal{V}_\ell$. Thus, we could not directly obtain the ventricular similarity. To do so, we need to know more about $\mathcal{E}_\ell$ so that we could design a way to obtain the ventricular similarity. 

Recall the physiological fact that while the ventricular response are similar from beat to beat dominated by the electrophysiological dynamics of the heart, they depend on, for example, the respiration and R peak to R peak interval; that is, while the ventricular response is a function and has many freedoms, it has several physiological constraints could be well approximated by a low dimensional geometric structure. Also recall the pathophysiological fact that the f-wave is somehow structured in the spectral domain, so the f-wave is not a totally random object and could be approximated by a low dimensional geometric object. These two facts lead us to consider the following {\em fiber bundle} assumption. 

\begin{Assumption}
The range of the random vector $\mathbf{E}:=\mathbf{V}+\mathbf{A}$ is a $D$-dim compact manifold. Moreover, this $D$-dim manifold is a fiber bundle, where the range of $\mathbf{V}$ form the base manifold and the range of $\mathbf{A}$ is diffeomorphic to the fiber. We assume that the dimension of the base manifold is much smaller than $D$.
\end{Assumption}

An equivalent way to view this model is that the $\ell$th surface ECG segments $\mathcal{E}_\ell$ are sampled from $\mathbf{V}$, with the contamination of the random vector $\mathbf{A}$. 
Note that this fiber bundle is trivial, and in this work we do not endow it with any further geometric structure. Its flexible structure will be explored in the future to include more physiological information. In practice, the  surface ECG signal is always contaminated by noise, and we have

\begin{equation}\label{ECGdecomp}
x_\ell(t)=E_\ell(t)+\xi(t)\,,
\end{equation}
where $x_\ell(t)$ is the recorded surface ECG signal and $\xi(t)$ is the noise. We should have the following assumption regarding the noise.

\begin{Assumption}
$\xi$ and $\mathbf{E}$ are independent.
\end{Assumption}

The above assumptions allow us to well group the surface ECG segments $\mathcal{X}_\ell=\{x_\ell|_{I_k}\}$ so that all cardiac activities in a clustered group have a similar ventricular activity.  
Specifically, based on the physiological knowledge, we could design a metric on $\mathcal{X}_\ell$ so that it is sensitive only to the VA but not to the f-wave. In our algorithm, the proposed ventricular similarity metric is the one for this purpose. To avoid the possible error and noise, we apply the DM to stabilize the ventricular similarity metric, which we call the diffusion ventricular similarity metric.

In sum, if we could successfully recover $v_\ell|_{I_k}$ from $\mathcal{X}$ for each $x_\ell|_{I_k}$, then we could get $a_\ell(t)+\xi(t)$ by a suitably chosen averaging process, like the non-local Euclidean median considered in this paper. Finally, via a denoise process to remove the noise $\xi(t)$, we could recover the atrial activity $a_\ell(t)$, and hence the f-wave.

\subsection{Diffusion map}\label{section:DM} 

Due to explosive advances in technology, it is a growing consensus that high-dimensional and massive data set analysis is the key to further understanding our universe. Usually, the collected dataset is represented as a {\em point cloud},
which is a set of data points $\mathcal{X} = \{x_i\}_{i = 1}^n$ in some high-dimensional space. To simplify the discussion, we assume that $\mathcal{X} \subset \mathbb{R}^p$, where $p$ might be huge. The ultimate goal is to organize the point cloud and extract useful information. To this goal, we need some relationship measurements between data points. Usually, the relationship, or often understood as the {\em affinity} or {\em similarity}, between data points is described by a \textit{kernel} function defined on the point cloud, $K \colon \mathcal{X} \times \mathcal{X} \rightarrow \mathbb{R}$, which is non-negative and symmetric and satisfies $K(x_i, x_i) = 1$ for every $i$. Kernels are often only defined or trusted locally, and they may be rough and distorted by noise. We wish to integrate local kernel values to obtain a metric defined on $\mathcal{X}$, which allows us to organize $\mathcal{X}$ and is robust to noise. 
In the past few decades, several algorithms have been proposed to define this kernel function, and apply some properties of the designed kernel function to organize the point cloud. We focus on the diffusion map (DM) \cite{Coifman_Lafon:2006} in this paper. It has been well-known that the DM is robust to noise \cite{ElKaroui:2010a,ElKaroui_Wu:2015b} and offers a robust metric on the point cloud called the diffusion distance (DD), which is able to faithfully reflect the underlying non-linear structure of the data set.

We now summarize the diffusion maps algorithm and the diffusion distance. 
We perform the integration of the local affinity information captured by the kernel function $K$ by introducing a Markov process that diffuses across $\mathcal{X}$.
First, define the symmetric affinity matrix $W\in \mathbb{R}^{n\times n}$ as 
\begin{equation}
W_{ij} = K(x_i, x_j)
\end{equation}
and the diagonal degree matrix $D\in \mathbb{R}^{n\times n}$ as
\begin{equation}
D_{ii} = \sum_{j = 1}^n K(x_i, x_j).
\end{equation}
The transition matrix $P$ which will govern our Markov process on the point cloud is defined as 
\begin{equation}
P = D^{-1}W.
\end{equation}
Due to the non-negative nature of the kernel function, we could interpret the value of $P_{ij}$ as being the probability of transitioning from $x_i$ to $x_j$ in one time step, and view the rows of $P$ as probability distributions on $\mathcal{X}$. Note that in practice, the kernel function could be truncated by removing entries smaller than a small chosen threshold. This leads to a sparse $P$, which not only preserves the geometric information but also reduces the computational load.

We have the following decomposition of $P$:
\begin{equation}
P = U \Lambda V^{-1},\quad 
U = 
\begin{bmatrix} 
\varphi_1 & \cdots & \varphi_n 
\end{bmatrix} 
,\quad
\Lambda = \textsf{diag} \left( \lambda_1,\,  \ldots,\, \lambda_n \right),
\quad
V = 
\begin{bmatrix} 
\phi_1 & \cdots & \phi_n 
\end{bmatrix} 
\end{equation} 
where $1 = \lambda_1 \geq \cdots \geq \lambda_n$ are the eigenvalues of $P$, and $\varphi_1,\cdots, \varphi_n$ (respectively $\phi_1,\cdots, \phi_n$) are the associated right (respectively left) eigenvectors of $P$. It has been well known that $U^\top V=V^\top U=I_{n\times n}$, which leads to $P^t = U \Lambda^t V^{-1}$, where $t>0$ and $P^t_{ij}$ describe the probability of diffusing from $x_i$ to $x_j$ in $t$ time steps. Note that we could find a complete eigenvectors of $P$ since $P$ is similar to the symmetric matrix $D^{-1/2}WD^{-1/2}$. 
The \textit{diffusion distance with time $t>0$} is then defined as 
\begin{equation}
d_{\textup{DD},t} \ \colon \mathcal{X} \times \mathcal{X} \rightarrow \mathbb R_{\geq 0},\quad 
d_{\textup{DD},t}({x}_i, {x}_j) = \Vert e^\top_i(P^t) - e^\top_j(P^t) \Vert_{L^2(\phi_1)},
\end{equation}
where $\phi_1$ is the stationary distribution of the Markov chain $P$, $\|v\|_{L^2(1/\phi_1)}:=\sqrt{\sum_{l=1}^nv(l)^2/\phi_1(l)}$ is the weighted $L^2$ norm on $\mathbb{R}^n$, and $e_i\in\mathbb{R}^{n}$ is a unit vector with the $i$th entry $1$.
The diffusion distance between $x_i$ and $x_j$ is the $\ell^2$ distance between the probability distributions associated to $x_i$ and $x_j$ after diffusing for time $t$.

In practice, due to the existence of noise, we may consider a truncation scheme to stabilize the diffusion distance.
Note that we have
\begin{equation}
\Vert e^\top_i(P^t) - e^\top_j(P^t) \Vert_{L^2(\phi_1)}=\Vert e^\top_i(U\Lambda^t) - e^\top_j(U\Lambda^t) \Vert_{L^2}\,,
\end{equation}
where the right hand side leads to the DM. Specifically, since $e^\top_i(U\Lambda^t)=(\lambda^t_1\varphi_1(i),\ldots,\lambda^t_n\varphi_n(i))^\top\in\mathbb{R}^n$, for $t>0$, we could define the DM with the diffusion time $t>0$ as
\begin{equation}
\Phi_t:x_i\to (\lambda^t_1\varphi_1(i),\ldots,\lambda_n^t\varphi_n(i))^\top\in\mathbb{R}^n.
\end{equation}
Clearly, if some eigenvalues are small, and the diffusion time is large, then we could ignore these eigenvalues and reduce the dimension of the embedded points, as well as preserve the DD. 
On the other hand, it has been shown in \cite{ElKaroui:2010a,ElKaroui_Wu:2015b} that the noise has a larger impact on the smaller eigenvalues, thus, it is encouraged to truncate the small eigenvalues. These thus lead to the following truncated DM and truncated DD:
\begin{equation}
\Phi^{(m)}_t:x_i\to (\lambda^t_1\varphi_1(i),\ldots,\lambda_{m+1}^t\varphi_{m+1}(i))^\top\in\mathbb{R}^{m+1},
\end{equation}
where $m\in\mathbb{N}$ is chosen by the user under some criteria.
We mention that if the graph associated with the kernel function is connected, then $\varphi_1$ is a constant vector and we could remove $\varphi_1$ in the embedding since it is not informative. 

The above algorithm could be carried out to any point cloud. However, in general, we have limited knowledge about the structure underlying the available point cloud and little could be inferred about the spectral structure. But it is widely accepted that a low-dimensional and possibly non-linear structure should exist. A common model to quantify this structure is assuming that $\mathcal{X}$ is sampled from a low-dimensional manifold. Under this assumption, we know that the DD approximates the geodesic distance locally \cite{Singer_Wu:2012}. Moreover, it has been well known that the eigenvalue decays exponentially fast (Weyl's law), so the truncation scheme could be chosen based on the knowledge of the manifold (see, for example, \cite[Appendix]{ElKaroui_Wu:2015b}) to achieve the stability property. More theoretical information about diffusion geometry can be found in \cite{Coifman_Lafon:2006,Singer_Wu:2016} and the reference therein.

The above theoretical properties of the DD justifies the accurate and robust metric design in our algorithm under the manifold setup. Specifically, the designed ventricular similarity metric is nothing but an approximation of the metric of the base manifold associated with the fiber bundle structure of the cardiac activity. Recall that the base manifold models the VA, while the fiber models the f-wave. In other words, the ventricular similarity metric is designed to ignore the fiber structure and focus on the metric on the base manifold. However, due to the possible noise incurred when we evaluate the ventricular similarity metric, we apply the DM to evaluate the diffusion ventricular similarity metric, which is more robust to the noise and could reflect the underlying ventricular similarity more accurately. A more theoretical justification of the fiber bundle analysis is out of the scope of this paper, and will be studied and reported in the future work. 

{\subsection{Visualization of ventricular activity and f-wave dataset} }

The existence of the low dimensional geometric structure is a critical assumption of the algorithm, which allows us to better organize the cardiac activity and hence determine the template for the VA. In this section, we show the underlying geometries of various real Holter signals. We use the DM and examine the embeddings given by the first three non-trivial eigenvectors.  We use the Gaussian kernel with bandwidth
\begin{equation}
2 \times \textsf{median}_i \{ \Vert \mathbf{x}_i - \mathbf{x}_{i,50} \Vert\}^2,
\end{equation}  
where $\mathbf{x}_{i,50}$ is the 50th neighbour of $\mathbf{x}_i$. We retain kernel information for each point's 200 nearest neighbours, and we throw away any outlying point whose distance to its 100th neighbour is less than 0.2. 

\begin{figure}
\centering
\includegraphics[width=.32\textwidth]{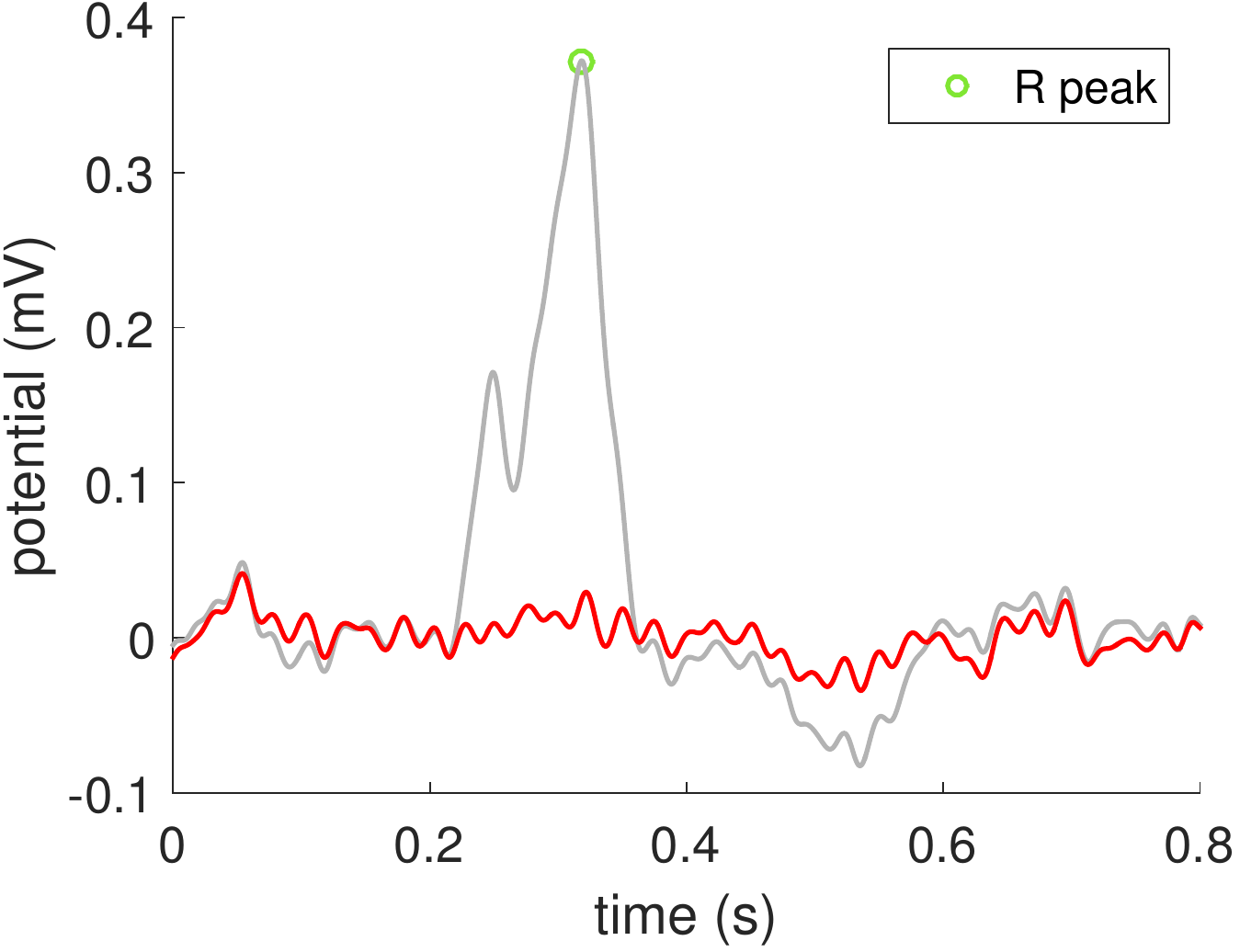}
\includegraphics[width=.32\textwidth]{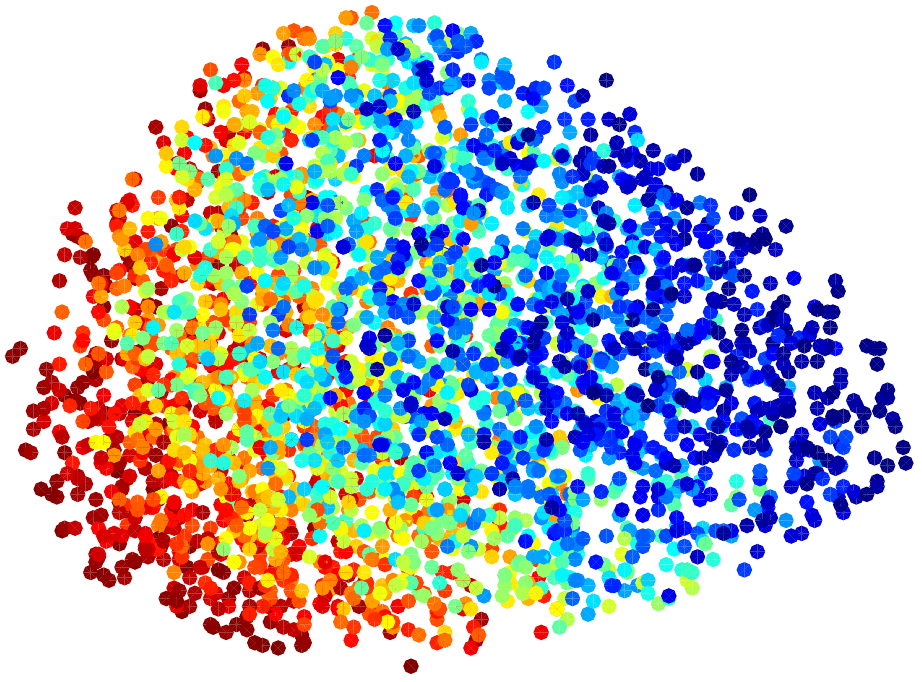}\\
\includegraphics[width=.32\textwidth]{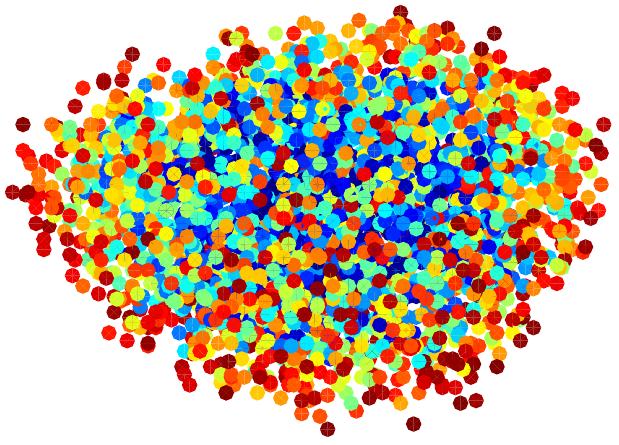}
\includegraphics[width=.32\textwidth]{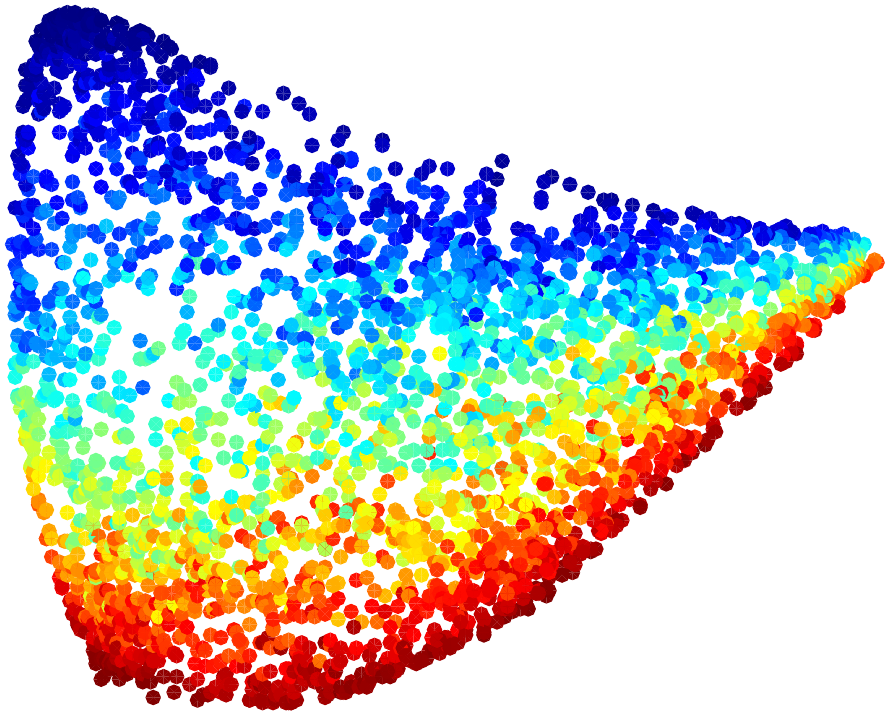}
\includegraphics[width=.32\textwidth]{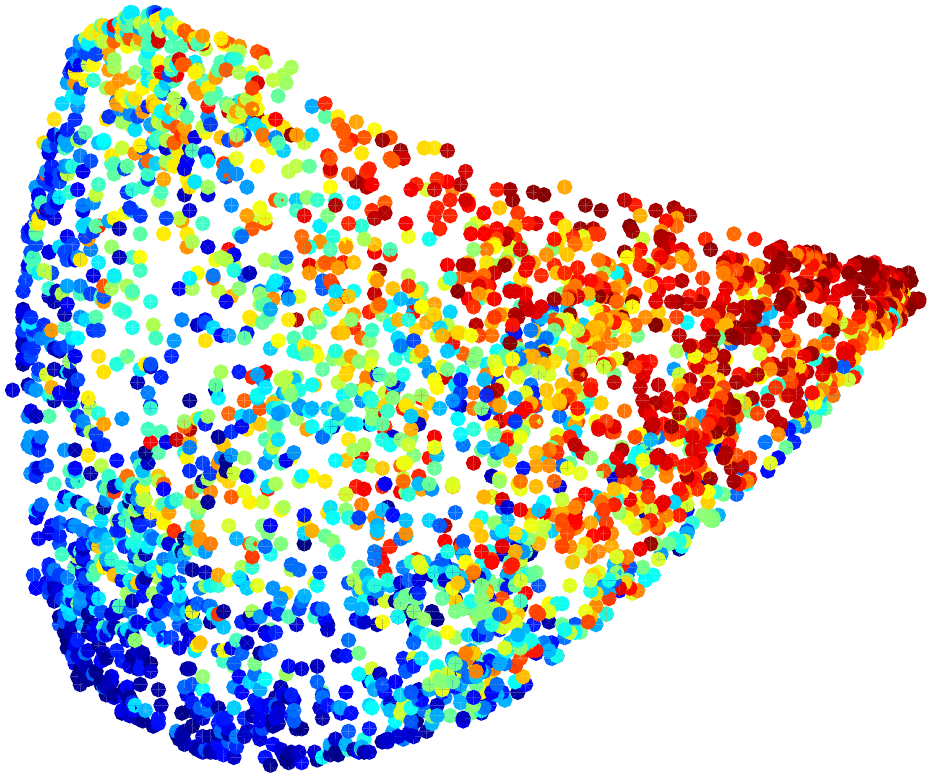}
\caption{Top left: Example patch $\mathbf{x}_i$ (grey) with the extracted f-wave $\tilde{\mathbf{a}}_i$ (red); top right: Embedding of patches $\{\mathbf{x}_i\}_{i=1}^{3347}$, where the colouring means the R peak height (red is high, blue is low); bottom left: Embedding of extracted f-wave segments $\{\tilde{\mathbf{a}}_i \}_{i=1}^{3347}$, where the colouring means the standard deviation (red is high, blue is low), which shows the fiber structure; bottom middle: Embedding of the extracted ventricular templates $\{ \tilde{\mathbf{v}}_i \}_{i=1}^{3347}$, where the colouring means the R peak height (red is high, blue is low); bottom right: Embedding of the extracted ventricular templates $\{ \tilde{\mathbf{v}}_i \}_{i=1}^{3347}$, where the colouring means the width of the QRS complex (red is high, blue is low). It is clear that the R peak height and QRS complex width well parameterized the ventricular templates.  \label{fwavenoise1}}
\end{figure}

Take a Holter signal whose f-wave behaves like noise. The signal contains 3347 beats.  Figure~\ref{fwavenoise1} shows the embedding of the patches space $\{\mathbf{x}_i\}_{i=1}^{3347}$, the extracted ventricular templates $\{\tilde{\mathbf{v}}_i\}_{i=1}^{3347}$, and the extracted f-waves $\{\tilde{\mathbf{a}}_i\}_{i=1}^{3347}$. It is clearly that the patch space $\{\mathbf{x}_i\}_{i=1}^{3347}$ could be parameterized by the R peak height, while the f-wave segments behave like ``noise'' and little structure could be observed. 
On the other hand, the embedding of the extracted ventricular templates $\{\tilde{\mathbf{v}}_i\}_{i=1}^{3347}$, is well parametrized by the R peak height. This result provides a partial evidence supporting our fiber bundle model for the patch space -- the f-wave could be modeled by a fiber, which adds dimensionality to the patch space. 

\begin{figure} 
\centering
\includegraphics[width=.32\textwidth]{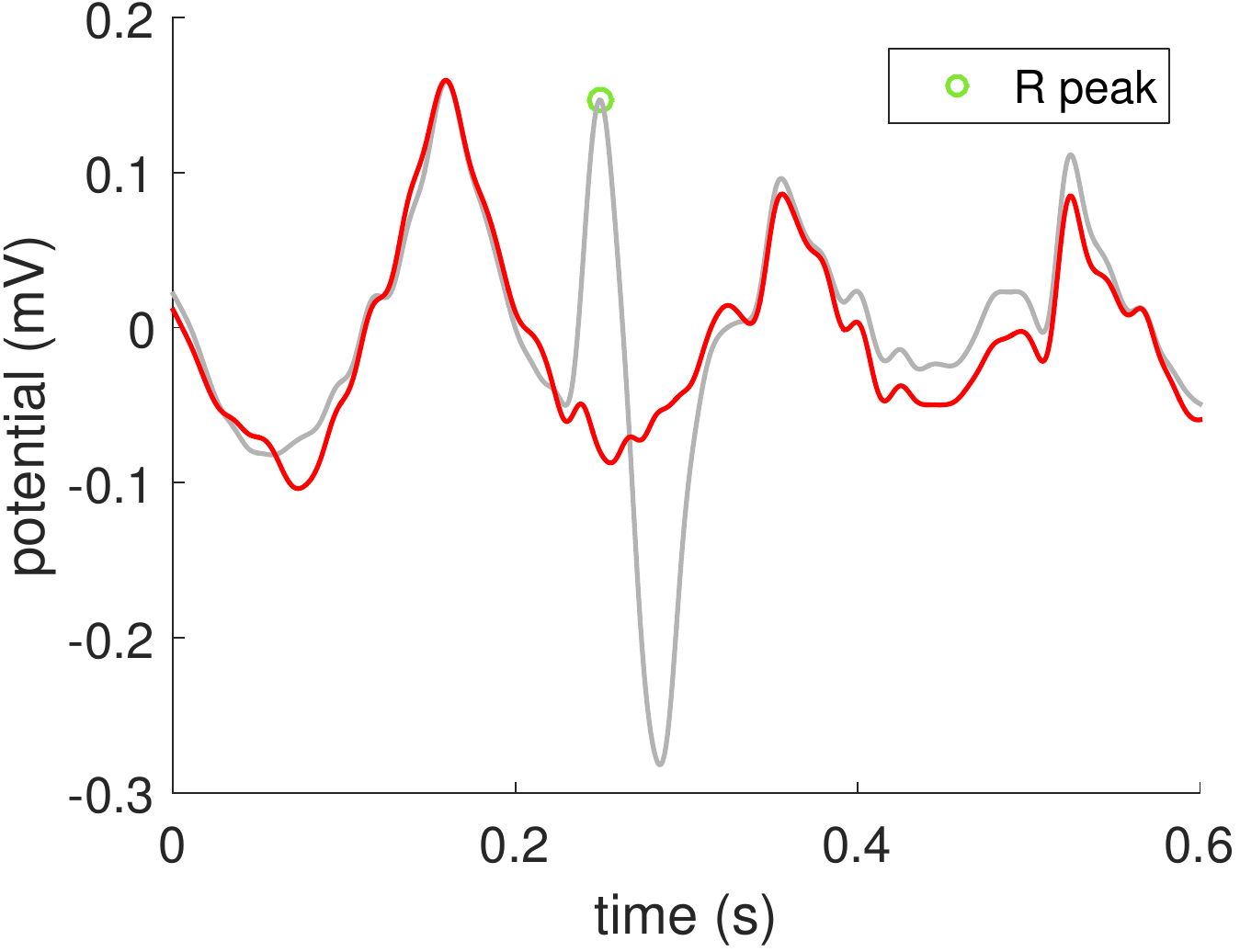}
\includegraphics[width=.32\textwidth]{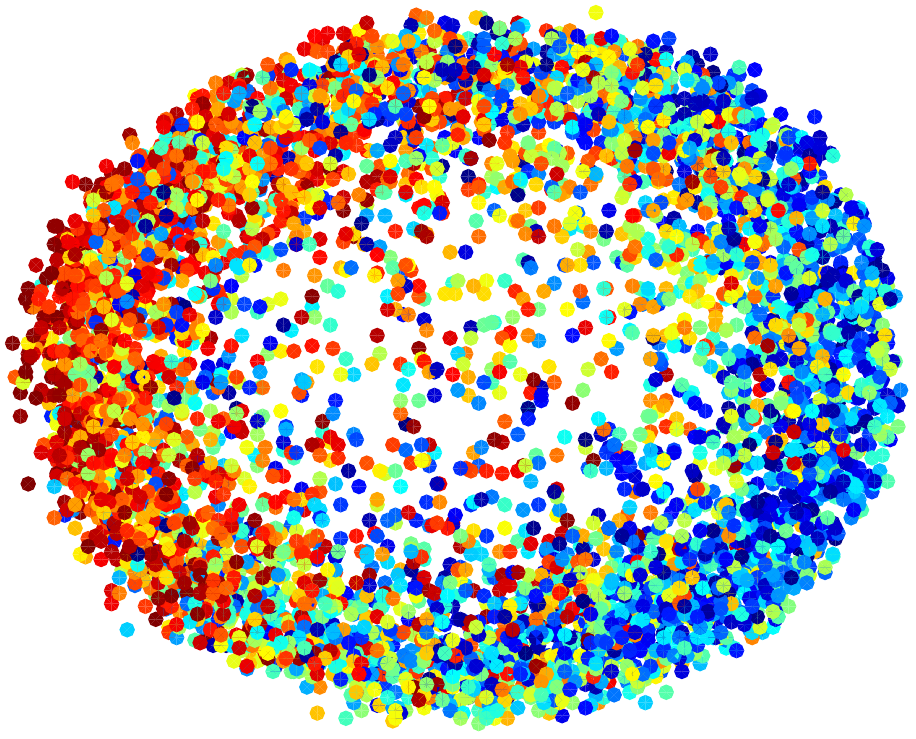}\\
\includegraphics[width=.32\textwidth]{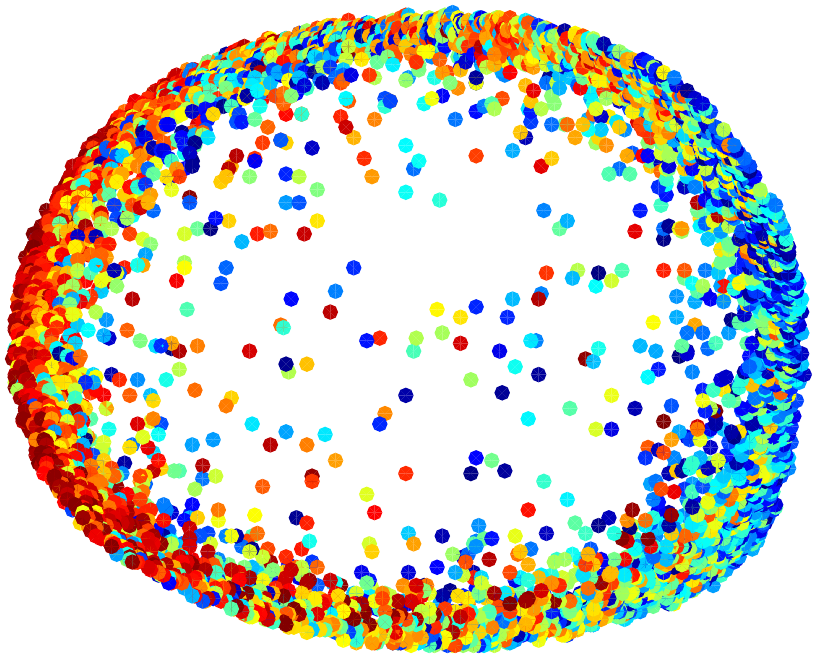}
\includegraphics[width=.32\textwidth]{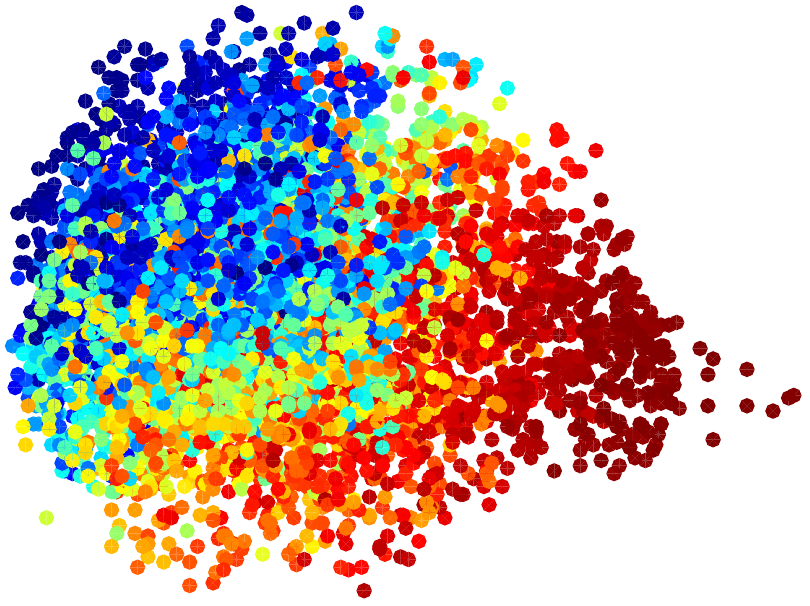}
\includegraphics[width=.32\textwidth]{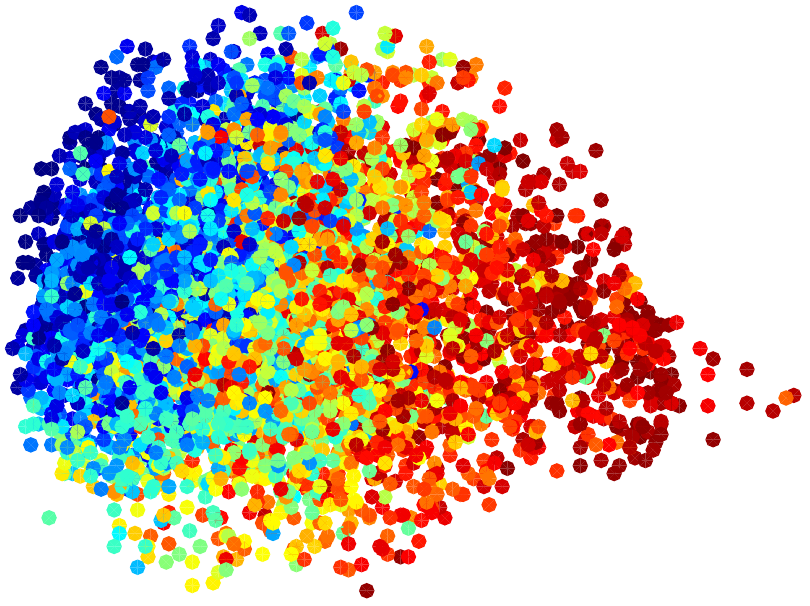}
\caption{Top left: Example patch $\mathbf{x}_i$ (grey) with extracted f-wave $\mathcal{A}:=\{\tilde{\mathbf{a}}_i\}_{i=1}^{5393}$ (red); top right: Embedding of patches $\mathcal{X}:=\{\mathbf{x}_i\}_{i=1}^{5953}$ showing a dominant f-wave, coloured by the first coordinate $\mathbf{x}_i(1)$; bottom left: Embedding of extracted f-wave segments $\mathcal{A}$, coloured by $\tilde{\mathbf{a}}_i(1)$; bottom middle:  Embedding of the extracted ventricular templates $\mathcal{V}:=\{ \tilde{\mathbf{v}}_i \}_{i=1}^{5953}$, coloured by the R peak height; bottom right: Embedding of the extracted ventricular templates $\mathcal{V}$, where the colouring means the width of the QRS complex (red is high, blue is low). Note that due to the fluttering nature, the f-wave is somehow ``periodic'', so the dataset $\mathcal{A}$ is distributed over a one-dimensional circle. Also, since the f-wave is dominant, the embedding of $\mathcal{X}$ is dominated by the embedding of $\mathcal{A}$.}\label{fwavedominant}
\end{figure}

To further evaluate how the f-wave behave as a fiber, we choose a Holter signal that has a strong and fluttering pattern. The one hour signal contains 5953 beats. Due to the fluttering nature, the f-wave is somehow ``periodic,'' so the fiber describing the f-wave should behave like a one-dimensional circle. To see this, we embed the corresponding patch space $\{\mathbf{x}_i\}_{i=1}^{5953}$ by the DM. In Figure~\ref{fwavedominant}, we see that compared with that in Figure \ref{fwavenoise1}, the DM of $\{\mathbf{x}_i\}_{i=1}^{5953}$ is close to the one-dimensional circle, and the coloring by the first coordinate of the extracted f-wave, $\tilde{\mathbf{a}}_i(1)$ further supports the periodicity behavior of the fluttering pattern. On the other hand, the ventricular template could not be easily seen because of the relative strength of the f-wave. Note that the noise in this particular embedding may be partially attributed to variations in ventricular activity.

\end{document}